\documentclass{IEEEoj}

\usepackage{amsmath,amssymb,amsfonts,amsthm,tabulary}
\usepackage{algorithmic, algorithm}
\usepackage{cite, amsfonts, amssymb, amsthm, bm, bbm, graphicx, relsize, multirow, booktabs, tikz,subfigure,soul}
\usepackage{cite}
\usepackage{amsmath,amssymb,amsfonts}
\usepackage{algorithmic}
\usepackage{graphicx,color}
\usepackage{textcomp}
\def\BibTeX{{\rm B\kern-.05em{\sc i\kern-.025em b}\kern-.08em
    T\kern-.1667em\lower.7ex\hbox{E}\kern-.125emX}}
\AtBeginDocument{\definecolor{ojcolor}{cmyk}{0.93,0.59,0.15,0.02}}
\def\OJlogo{\vspace{-4pt}\hskip-4pt\includegraphics[height=18pt]{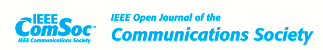}}

\newcommand{\beq}{\begin{equation}}
\newcommand{\eeq}{\end{equation}}

\newcommand{\bA}{\mathbf{A}}
\newcommand{\bB}{\mathbf{B}}

\newcommand{\ds}{\displaystyle}
\newcommand{\norm}[1]{\left\lVert#1\right\rVert} 

\begin{document}
\receiveddate{XX Month, XXXX}
\reviseddate{XX Month, XXXX}
\accepteddate{XX Month, XXXX}
\publisheddate{XX Month, XXXX}
\currentdate{11 January, 2024}
\doiinfo{OJCOMS.2024.011100}

\title{Co-existing/Cooperating  Multicell Massive MIMO and  Cell-Free Massive MIMO Deployments: Heuristic Designs and Performance Analysis}

\author{STEFANO BUZZI\IEEEauthorrefmark{1,2,3} \IEEEmembership{(Senior Member, IEEE)}, CARMEN D'ANDREA\IEEEauthorrefmark{1,3} \IEEEmembership{(Member, IEEE)}, 
LI WANG\IEEEauthorrefmark{4},
        AHMET HASIM GOKCEOGLU\IEEEauthorrefmark{4},
        AND~GUNNAR~PETERS\IEEEauthorrefmark{4}
}
\affil{Department
	of Electrical and Information Engineering, University of Cassino and Southern Latium, I-03043 Cassino (FR), Italy}
\affil{Department of Electronics, Information and Bioengineering, Politecnico di Milano, Milan, Italy}
\affil{Consorzio Nazionale Interuniversitario per le Telecomunicazioni (CNIT), Parma, Italy}
\affil{Huawei's Sweden Research Center, Stockholm, Sweden}
\corresp{CORRESPONDING AUTHOR: Stefano Buzzi (e-mail: buzzi@unicas.it).}
\authornote{This work has been supported by Huawei Technologies Sweden AB. \\
S. Buzzi and C. D'Andrea have also been supported by the European Union under the Italian National Recovery and Resilience Plan (NRRP) of NextGenerationEU, partnership on “Telecommunications of the Future” (PE00000001 - program “RESTART”, Structural Project 6GWINET and Spoke 3 Cascade Call Project SPARKS}
\markboth{Preparation of Papers for IEEE OPEN JOURNALS}{Author \textit{et al.}}

\begin{abstract}
Cell-free massive MIMO (CF-mMIMO) systems represent a deeply investigated  evolution from the conventional multicell co-located massive MIMO (MC-mMIMO) network deployments. Anticipating a gradual integration of CF-mMIMO systems alongside pre-existing MC-mMIMO network elements, this paper considers a scenario where both deployments coexist, in order to serve a large number of users using a shared set of frequencies. The investigation explores the impact of this co-existence on the network's downlink performance, considering various degrees of mutual cooperation, precoder selection, and power control strategies.
Moreover, to take into account the effect of the proposed cooperation scenarios on the fronthaul links, this paper also provides a fronthaul-aware heuristic association algorithm between users and network elements, which allows the fulfillment of the front-haul requirement on each link. The research is finally completed by extensive simulations, shedding light on the performance outcomes associated with the various levels of cooperation and several solutions delineated in the paper.
\end{abstract}


\maketitle

\section{INTRODUCTION}
\IEEEPARstart{C}{ell}-free massive MIMO (CF-mMIMO) \cite{demir2021foundations}  is one of the key technologies considered for the evolution of traditional multicell massive MIMO (MC-mMIMO) architectures. CF-mMIMO was proposed in 2015
\cite{ngo2015cell} as a way to overcome the QoS disparity experienced by active users in MC-mMIMO networks using co-located large antenna arrays. In such systems, indeed, users located at the center of the cell in close proximity to the base station (BS)  antenna arrays can enjoy a much better performance than users located at the cell-edge, due to the interference coming from adjacent cell base stations. In a CF-mMIMO system, instead, the BSs using co-located antenna arrays are substituted by a large number of simpler and cheaper access points (APs), equipped with a small number of antennas, and connected through a central processing unit (CPU) through a high-speed fronthaul link. In the user-centric implementation of CF-mMIMO, each user is served by a limited number of surrounding APs \cite{buzzi_CFUC2017,buzzi2017user}, and each AP usually serves a limited number of users. CF-mMIMO systems have been shown to outperform traditional MC-mMIMO systems in terms of fairness between users; moreover, when used at mmWave carrier frequencies, they provide macro-diversity and protection against signal blockages \cite{Alonzo_Buzzi_TGCN2019}, while in low-load situations they are more energy efficient than MC-mMIMO deployments \cite{Ngo_EE_cellfree}. 

\subsection{Related work}
CF-mMIMO deployments have attracted, since their introduction, a great deal of interest, both in academia and in industry, and several research groups worldwide have been investigating their suitability as a possible evolution of legacy massive MIMO systems. The work \cite{buzzi2017user} introduced the paradigm shift from non-scalable cell-free systems to user-centric communications, where each user is served by a limited number of surrounding APs. The paper \cite{Ngo_CellFree2017}
explores the trade-offs and advantages of CF-mMIMO in contrast to small cell deployments, showing that this deployment is capable of achieving greater user fairness than a small-cell network architecture. In \cite{bjornson2019CF_MMSE}, 
the authors investigate methods to make cell-free massive MIMO systems competitive with Minimum Mean Squared Error (MMSE) processing, advocating the advantages of centralized processing of the baseband data collected in APs, as opposed to more decentralized implementations where some data processing and beamforming computation is executed locally in each AP.   The pursuit of scalability in CF-mMIMO systems was addressed, besides the papers \cite{buzzi_CFUC2017,buzzi2017user}, also in \cite{bjornson2020scalable}, where a formal definition of what a scalable CF-mMIMO system is was introduced, along with practical algorithms for associating APs to the UEs and for assigning pilots to the UEs. The problem of scalability was also simultaneously addressed  in \cite{Interdonato_Scalability2019}, where solutions similar to those in \cite{bjornson2020scalable} were outlined.  

The paper \cite{chen2020structured} focuses  on structured massive access in CF-mMIMO systems with a very large number of users, focusing on the issues of AP-UE association, pilot allocation to UEs, and analysis of different local combining rules at the APs for uplink decoding. Effective pilot allocation is also addressed in \cite{buzzi2020pilot} , where a solution based on the Hungarian algorithm is proposed. In \cite{papazafeiropoulos2020performance}, the authors use stochastic geometry tools to contrast CF-mMIMO systems to small cell deployments; in particular, assuming that the APs are distributed according to a Poisson point process, closed form expressions for the downlink spectral efficiency and coverage probability, showing that CF-mMIMO outperform small-cells systems in both performance measures. 
In \cite{d2020analysis}, the suitability of CF-mMIMO to support communications with unmanned aerial vehicles (UAVs) is investigated and a comparison with the performance attained by a MC-mMIMO deployment is offered. A similar topic is addressed in the more recent work \cite{elwekeil2022power}, in which power control algorithms aimed at optimizing the users' rate in the finite-blocklength regime are developed using successive convex optimization. In \cite{mukherjee2020edge,interdonato2023joint}, CF-mMIMO systems are examined in which APs also have computing capabilities to help UEs perform intense computational tasks. In general, the papers prove that the CF-mMIMO network architecture facilitates mobile edge computing implementations, since it helps to bring network computational power as close as possible to the end users. 
{The suitability of CF-mMIMO systems to enable ultra-reliable and low-latency communications (URLLC) is studied in the reference \cite{Zhang_URLLC2024}. By utilizing short-packet transmission and accounting for imperfect channel state information, the authors propose a path-following power control algorithm, maximizing the downlink sum-rate, designed to meet the stringent latency and reliability requirements of URLLC. In reference \cite{ZhangRSMA2024}, then, secure transmission in a downlink rate splitting multiple access (RSMA) CF-mMIMO system is addressed by incorporating in the analysis the low-resolution digital-to-analog converters.} 
Very recently, CF-mMIMO implementations based on the O-RAN architecture have also started being investigated; in particular, in  \cite{beerten2022user} a study is performed about the CF-mMIMO functionalities that can be implemented using the Near-Real Time RAN Intelligent Controller (Near-RT RIC). In the paper \cite{oh2023decentralized}, instead, the O-RAN architecture is leveraged to implement a pilot assignment scheme using multi-agent deep reinforcement learning. These studies give evidence of how the scientific community is focused on addressing practical problems to enable the real deployment of CF-mMIMO systems in the near future. 

One drawback/bottleneck of CF-mMIMO systems is the data rate required on the fronthaul connection between the APs and the CPU. On the downlink, data symbols to be sent to the users and (possibly) beamforming coefficients are sent to the APs from the CPU, whereas, on the uplink, each AP sends, for each of its connected users, a complex sufficient statistic in each data symbol to enable data detection(decoding at the CPU. The fronthaul rate is thus a critical element of a CF-mMIMO system, and the problem of a fronthaul-aware implementation of CF-mMIMO has received considerable attention. In particular, in \cite{femenias2019}, with reference to a CF-mMIMO system operating at mmWave carrier frequencies, power allocation and fronthaul quantization optimization schemes are derived promoting system fairness across users and subject to a fronthaul capacity constraint. In \cite{femenias2020fronthaul}, analytical closed-form expressions for achievable user rates in both the uplink (UL) and downlink (DL) of a fronthaul-capacity constrained cell-free massive MIMO network employing low-resolution ADCs are derived, and  max-min fairness power control optimization problems are considered, using the Bussgang decomposition for the output of finite-resolution ADCs. 
The paper \cite{masoumi2019performance} investigates the uplink of a CF-mMIMO system with limited fronthaul capacity, considering hardware impairments. Three different transmission strategies with different amounts of local processing at the APs before quantization and uplink transmission on the fronthaul are analyzed and compared. Finally, the paper
\cite{bashar2020exploiting} analyzes the case where both the channel estimates and the signals received at the APs are quantized and sent to the CPU, and the case where local signal combination takes place at the APs before quantization and transmission at the CPU. For both cases, the sum-rate maximization problem is considered, and, exploiting deep learning, a mapping is learned from the large-scale fading coefficients to the optimal power allocation through solving the sum-rate maximization problem using the quantized channel.

It should be noted that all of the above-cited papers, similar to the vast majority of papers on CF-mMIMO systems, assume that these systems operate in dedicated frequency bands and are not interfered by other wireless systems. Otherwise stated, the open literature seems to have overlooked the relevant problem of the initial rollout of CF-mMIMO systems, which presumably will happen using frequency bands used by other network architectures such as MC-mMIMO systems.
This is, for instance, the view expressed in \cite{Kim_Magazine2022}, which discusses a transitional system
architecture where the cell-free
system and the legacy cellular system coexist, and in \cite{Interdependent_CF_Cellular2023}, where the high costs associated with the deployment and management of a large number of APs over a large area are presented as a motivation to consider a gradual deployment. These references, i.e. \cite{Kim_Magazine2022, Interdependent_CF_Cellular2023} are magazine articles discussing the integration of CF-mMIMO and MC-mMIMO at a high level, and with a special focus on the challenges posed by the integration at the higher layers of the network protocol stack. 
In contrast, papers \cite{rezaei2020underlaid,galappaththige2021exploiting} focus on the specific physical layer and MAC designs of such integration. Specifically, in the former paper,  
the achievable rate of a hybrid CF-mMIMO nonorthogonal multiple-access (NOMA) system operating under a primary massive MIMO system was investigated, and a closed-form sum rate expression is derived for Rayleigh fading channels. The results indicate that the NOMA-based underlay CF-mMIMO efficiently utilizes scarce spectrum bands compared to orthogonal multiple-access. In the latter paper \cite{galappaththige2021exploiting}, instead, the authors investigate underlay spectrum sharing in CF-mMIMO systems, where a primary CF-mMIMO system serves primary users, and a secondary CF-mMIMO system exploits underlay spectrum sharing to serve secondary users. Rigorous power constraints and user-centric clustering are employed, achieving enhanced performance through reduced transmit powers and macro-diversity gains.

This paper focuses on this line of research providing further results and contributions related to the cooperation and co-existence of MC-mMIMO and CF-mMIMO systems, as detailed in the next section. 

\subsection{Paper contribution}

This paper considers a scenario where a CF-mMIMO system is deployed in the same area where a MC-mMIMO deployment already exists. {As already commented, the rationale for exploring this scenario stems from the gradual integration of CF-mMIMO network elements by operators into their systems, serving as both an enhancement and a complement to current cellular infrastructures.} Differently from \cite{rezaei2020underlaid}, it is here assumed that there is no hierarchical ordering between the CF-mMIMO system and the MC-mMIMO system, since both systems belong to and are controlled by the same operator; similarly, the users pool is unique and there is no distinction between users that are to be served by one system only and users that are to be served by the other system only. Likewise, the scenario considered here is different from that in \cite{galappaththige2021exploiting}, where two competing CF-mMIMO systems insist on the same set of carrier frequencies to serve the respective users.  
Several levels of cooperation between the two deployments are proposed and examined, under several configurations of the beamformers, the association rules between UEs and BSs and APs, and power control rules. 
Specifically, we assess the effect on the system performance of the following four configurations:
\begin{itemize}
	\item[a)] A deployment with MC-mMIMO BSs, arranged over equally spaced sites with 120 deg sectors, and with no-BS cooperation. 
	\item[b)] An heterogeneous non cooperative network where a MC-mMIMO and a CF-mMIMO system co-exist; it is assumed here that each UE can connect either to one AP or to one BS, and that there is no cooperation.
	\item[c)] A MC-mMIMO network co-existing with a CF-mMIMO network with horizontal cooperation; it is assumed here that each UE can be cooperatively served by either a certain number of BSs or by a certain number of APs. 
	\item[d)] A MC-mMIMO network co-existing with a CF-mMIMO network with full cooperation; it is assumed here that all the UEs are cooperatively and jointly served by several APs and several BSs. 
\end{itemize}
The above configurations are studied on the downlink for several types of precoding schemes, power control algorithms, and deployment strategies for cell-free APs.

Moreover, the fronthaul requirement needed to implement the several cooperative schemes is analyzed, and a practical design for the composition of the cooperating clusters is proposed, which tightly fulfills the maximum allowed fronthaul data rate. 
Differently from other papers dealing with fronthaul constraints, in this paper no quantization of the observable is used, and the obtained results do not rely on the Bussgang decomposition and on the random modeling of the quantization noise. Rather, the paper proposes a simple control algorithm that deterministically ensures that the fronthaul constraint is fulfilled by limiting the number of UEs that can be connected to each AP. 

In general, simulation results will highlight the benefits of increased integration between APs and BSs, revealing notable performance gaps between peripheral (cell-edge) and proximate users to BSs. Collaborative efforts significantly benefit peripheral users compared to those closer to BSs. Additionally, results will evidence instances where FH data rate limitations could counteract centralized beamforming advantages, as well as that in fully cooperative network setups implementing power control strategies to improve fairness among users may not always be optimal.

The sequel of this paper is organized as follows. The next section contains the description of the system model; it reports the full details on the four different analyzed cooperation deployment, on the channel model, the uplink channel estimation procedure, and also provides a pilot assignment algorithm for the UEs, and the expression of the downlink Signal-to-Interference plus Noise Ratio (SINR). Section \ref{DL_SE_Bounds} contains a discussion of the spectral efficiency bounds used on downlink. Section \ref{System_design} tackles the problem of network design and illustrates the basic association rules between UEs and APs/BSs, the downlink beamforming strategy, and the considered power allocation. Section \ref{Joint_PZF_Section} delves into the fully cooperative APs/BSs scenario and provides a practical procedure to design centralized beamformers while adhering to individual power constraints at the APs and at the BSs. Section \ref{Fronthaul_Section} is still devoted to the fully cooperative scenario and proposes the FH-compliant association algorithm beween the UEs and the APs/BSs. In section \ref{Performance_Evaluation} the extensive numerical results are provided and commented, while, finally, concluding remarks are given in Section \ref{Conslucions}.

\paragraph*{Notation}
In the following, lowercase and uppercase nonbold letters are used for scalars, $a$ and $A$, lowercase boldface letters, $\mathbf{a}$, for vectors and uppercase boldface letters, $\mathbf{A}$, for matrices. The transpose, the inverse and the conjugate transpose of a matrix $\mathbf{A}$ are denoted by $\mathbf{A}^T$, $\mathbf{A}^{-1}$ and $\mathbf{A}^H$, respectively. The generic entry $(m,n)$ of the matrix $\mathbf{A}$ is denoted as $\mathbf{A}[m,n]$, the $m$-th row is denoted as $\mathbf{A}[m,:]$ and the $n$-th column as $\mathbf{A}[:,n]$. The selection of the submatrix of $\mathbf{A}$ containing the rows from the $m$-th to the $\ell$-th and the columns from the $n$-th to the $q$-th is denoted as $\mathbf{A}[m:\ell,n:q]$. The diagonal matrix obtained by the scalars $a_1,\ldots, a_N$ is denoted by diag$( a_1,\ldots, a_N)$. The $N$-dimensional identity matrix is denoted as $\mathbf{I}_N$. The real and imaginary parts of a complex vector $\mathbf{z}$ are denoted as $\text{Re}\lbrace \mathbf{z} \rbrace$ and $\text{Im}\lbrace \mathbf{z} \rbrace$, respectively. The complex circularly symmetric Gaussian random variable with mean $\mu$ and variance $\sigma^2$ is denoted as $\mathcal{CN}\left(\mu,\sigma^2\right)$ and the statistical expectation operator is denoted by $\mathbb{E} \left[ \cdot \right]$.

\section{System model} \label{System_model}

We consider a network that consists of multi-antenna APs and macro-BSs, and single-antenna MSs. 
The APs are connected by means of fronthaul links to a CPU where data encoding/decoding is performed. Depending on the type of cooperation envisaged between the BSs and the APs, the BSs can also be connected through fronthaul links to a CPU. All communications take place on the same frequency band, i.e. uplink and downlink are separated through time-division-duplex (TDD), and there is perfect synchronization in the system. 

Although the MSs are equipped with a single antenna, each AP is equipped with a uniform linear array (ULA) made of $N_{\rm AP}$ antennas with half-wavelength spacing. We assume that BS sites are actually made of three sectors and the sectors are treated as independent base stations, i.e., we have $L$ BSs arranged in $L/3$ BS sites. Each sector is equipped with a ULA made of $N_{\rm BS}$ antennas with half-wavelength spacing.

 {In Table \ref{table:definition_symb} we report the definition of the main symbols used in the document.}

\begin{table}[]
	\centering
	\caption{ {Definition of the main symbols used in the paper}}
	\label{table:definition_symb}
	\def\arraystretch{1.2}
	\begin{tabulary}{\columnwidth}{ |p{1cm}|p{6.5cm}| }
		\hline
		\textbf{Name} 				& \textbf{Meaning} \\ \hline
		$L$ 				& number of BSs (sectors)  \\ \hline
  $M$ 				& number of APs  \\ \hline
  $K$ 				& number of users \\ \hline
		$N_{\rm BS}$ 				& number of antennas at the BSs\\ \hline
		$N_{\rm AP}$ 				& number of antennas at the APs\\ \hline
		$\mathbf{h}_{k,\ell}$				&  channel between the $k$-th user and the $\ell$-th BS \\ \hline
		$\mathbf{g}_{k,m}$			&  channel between the $k$-th user and the $m$-th AP \\ \hline
  $\rho_{k,\ell}$ &  scalar coefficients modeling the channel path loss and shadowing effects between the $k$-th UE and the $\ell$-th BS\\ \hline
   $\beta_{k,m}$ & scalar coefficients modeling the channel path loss and shadowing effects between the $k$-th UE and the $m$-th AP\\ \hline 
   $\tau_p$ & length of the training phase in samples \\ \hline
   $\boldsymbol{\phi}_k$ & pilot sequence transmitted by the $k$-th user \\ \hline
      $\eta_k$ & power transmitted by the $k$-th user during the training  phase\\ \hline
      		$\widehat{\mathbf{h}}_{k,\ell}$				&  estimate of the channel between the $k$-th user and the $\ell$-th BS \\ \hline
		$\widehat{\mathbf{g}}_{k,m}$			&  estimate of the channel between the $k$-th user and the $m$-th AP \\ \hline
     $b_{k,\ell}$ & the binary variates which is 1 if the $k$-th user is served by the $\ell$-th BS and 0 otherwise \\ \hline
          $a_{k,m}$ & the binary variates which is 1 if the $k$-th user is served by the $m$-th AP and 0 otherwise \\ \hline
          $\mathbf{w}_{k,\ell}^{(b)}$ & unit-norm  beamforming vector  used at the $\ell$-th BS for the downlink transmission to the $k$-th user \\ \hline
           $\mathbf{w}_{k,m}^{(a)}$ & unit-norm  beamforming vector  used at the $m$-th AP for the downlink transmission to the $k$-th user \\ \hline
          $\eta_{k,\ell}^{(b)}$ & positive scalar coefficient controlling the power transmitted by the $\ell$-th BS to the $k$-th user \\ \hline
          $\eta_{k,m}^{(a)}$ & positive scalar coefficient controlling the power transmitted by the $m$-th AP to the $k$-th user \\ \hline
		$P_{\ell,{\rm max}}^{(b)}$	&   maximum power available at the $\ell$-th BS \\ \hline
		$P_{m,{\rm max}}^{(a)}$	&   maximum power available at the $m$-th AP \\ \hline
		
	\end{tabulary}
\end{table}

\subsection{Network deployment scenarios} 
\begin{figure*}[]
	\centering
	\includegraphics[scale=0.58]{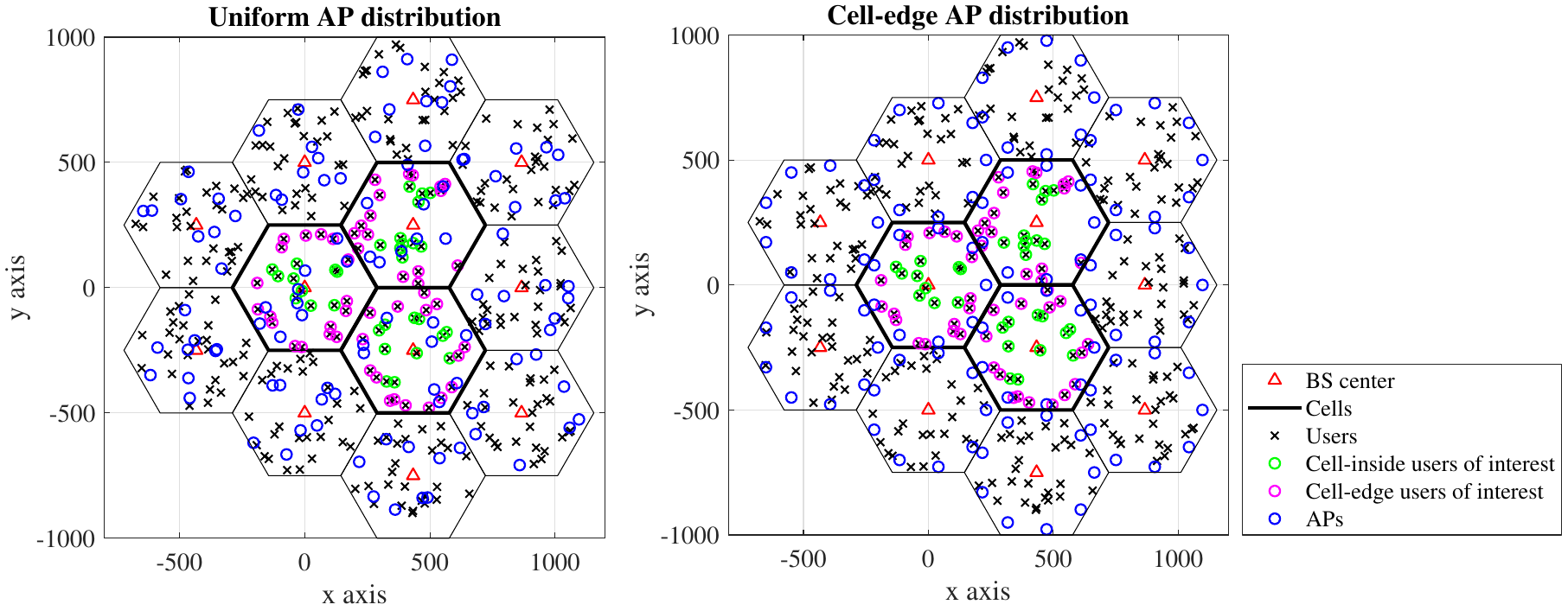}
	\caption{The simulated network deployment. The performance results that will be reported will refer only to the three cells with bolder edges. The remaining surrounding cells are considered to avoid border effects and take into account the extra-cell interference.}
	\label{Fig:scenario}
\end{figure*}

Although the subsequent derivations hold for any arbitrary configuration for the deployment of BSs and of the APs, our analysis will refer to a network deployment in which the three-sector macro BS sites are placed according to the classical hexagonal layout, while APs are deployed according to two possible strategies, namely uniform deployment or cell-edge deployment. The former case accounts for the situation where we want to densify the network by bringing network terminating points closer to the users so as to have the capability to further push the system capacity; the latter deployment, instead, aims similarly at increasing the network throughput, but exploiting APs to provide better coverage at the critical cell-edge, so as to increase the system fairness across users.  Figure \ref{Fig:scenario} shows an example of the two deployments considered. In particular, in the figure we also highlight users that are at a distance from a BS that is smaller than one third of the inter-site distance (ISD) and the remaining users. The former are called cell-inside users, while the latter will be termed cell-edge users. This distinction is made to study the effects on the two sets of the network deployment schemes considered below. 

As already discussed, this paper considers four possible network deployments, that are here summarized again
\begin{itemize}
	\item[a)] A deployment with $L$ MC-mMIMO BSs, arranged over equally spaced sites with 120 deg sectors, and with no-BS cooperation. This scenario will be indicated as ``MC-mMIMO''.
	\item[b)] An heterogeneous non-cooperative network where a MC-mMIMO and a CF-mMIMO system co-exist; it is assumed here that each UE can connect either to one AP or to one BS, and that there is no cooperation. This scenario resembles a network with base stations of two different kinds and will be named ``HET-noCoop''. 
	\item[c)] A MC-mMIMO network co-existing with a CF-mMIMO network with horizontal cooperation; it is assumed here that each UE can be cooperatively served by either a certain number of BSs or by a certain number of APs. However, there is no joint cooperation among BSs and APs. This scenario will be indicated as ``HORIZONTAL-Coop''. 
	\item[d)] A MC-mMIMO network co-existing with a CF-mMIMO network with full cooperation; it is assumed here that all the UEs are cooperatively and jointly served by several APs and several BSs. This is a fully cooperative scenario and will be labeled as ``FULL-Coop''.
\end{itemize}

\subsection{Channel Model} \label{Channel_model}
{The channel model under consideration in the frequency domain includes a combination of a large-scale fading coefficient (covering shadowing and path-loss effects) and a small-scale fading vector. We generally assume that the latter follows a Ricean distribution to represent the potential existence of a line-of-sight (LOS) component over short-distance links. \cite{d2020analysis}.
More precisely, $\mathbf{g}_{k,m} \in \mathbb{C}^{N_{\rm AP}}$ ($\mathbf{h}_{k,\ell} \in \mathbb{C}^{N_{\rm BS}}$) denotes the channel between the $k$-th user and the $m$-th AP ($\ell$-th BS), and is expressed as
\begin{equation}
\mathbf{g}_{k,m}=\sqrt{\frac{\beta_{k,m}}{K_{k,m}^{(a)}+1}} \left[ \sqrt{K_{k,m}^{(a)}} e^{j \vartheta_{k,m}} \mathbf{a}_{AP}\left(\theta_{k,m}\right) + \overline{\mathbf{g}}_{k,m}  \right] \; ,
\label{channel_model_a}
\end{equation}
\begin{equation}
\mathbf{h}_{k,\ell}=\sqrt{\frac{\rho_{k,\ell}}{K_{k,\ell}^{(b)}+1}} \left[ \sqrt{K_{k,\ell}^{(b)}} e^{j \varphi_{k,\ell}} \mathbf{a}_{BS}\left(\phi_{k,\ell}\right) + \overline{\mathbf{h}}_{k,\ell}  \right] \; .
\label{channel_model_b}
\end{equation}
In the above equations, 
$\beta_{k,m}$ ($\rho_{k,\ell}$) is a scalar coefficient modeling the large-scale fading, i.e.  the channel path loss and shadowing effects, between the $k$-th UE and the $m$-th AP ($\ell$-th BS). Moreover, 
 $\overline{\mathbf{g}}_{k,m}$ ($\overline{\mathbf{h}}_{k,\ell}$) is an $N_{\rm AP}$-dimensional vector ($N_{\rm BS}$-dimensional vector), whose entries are i.i.d. ${\cal CN}(0,1)$ random variates (RVs) modeling the fast fading. The parameter $K_{k,m}^{(a)}$ $\left(K_{k,\ell}^{(b)}\right)$ is the Rician $K$-factor for the channel from the $k$-th user to the $m$-th AP  ($\ell$-th BS); this factor usually depends on the probability that there is a LOS link between the user and the AP (BS). Letting $p_{\rm LOS}^{(a)}\left(d_{k,m}^{(a)}\right)$ $\left(p_{\rm LOS}^{(b)}\left(d_{k,\ell}^{(b)}\right)\right)$ denote the probability that there is a LOS link between the $k$-th user and the $m$-th AP ($\ell$-th BS),  the Ricean factors are evaluated as~\cite{d2020analysis},
  \begin{equation}
     K_{k,m}^{(a)}= \frac{p_{\rm LOS}^{(a)}\left(d_{k,m}^{(a)}\right)}{ 1 -  p_{\rm LOS}^{(a)}\left(d_{k,m}^{(a)}\right)}\, ,
 \end{equation}
 and
 \begin{equation}
     K_{k,\ell}^{(b)}= \frac{p_{\rm LOS}^{(b)}\left(d_{k,\ell}^{(b)}\right)}{ 1 -  p_{\rm LOS}^{(b)}\left(d_{k,\ell}^{(b)}\right)}\, ,
 \end{equation}
 respectively.
The quantities $\vartheta_{k,m}$ and $\varphi_{k,\ell}$ in \eqref{channel_model_a} and \eqref{channel_model_b} follow a uniform distribution in $[0, 2\pi]$, and denote the random phase offsets associated to the direct path from the $k$-th user to the $m$-th AP and to the $\ell$-th BS, respectively. Finally, $\mathbf{a}_{AP}\left(\theta_{k,m}\right) \in \mathbb{C}^{N_{\rm AP}}$ and $\mathbf{a}_{BS}\left(\phi_{k,\ell}\right) \in \mathbb{C}^{N_{\rm BS}}$ represent the steering vectors characterizing the direct path between the $k$-th user and the $m$-th AP or the $\ell$-th BS, respectively, and follow the model referenced in ~\cite{d2020analysis}. 
More details on the channel model will be given in Section \ref{Performance_Evaluation}.\ref{Simulation_parameters_section} and in the Appendix.}

\subsection{Uplink Training} \label{LMMSE_channel_estimation}

The dimension in time/frequency samples of the channel coherence length is denoted by $\tau_c$, and the dimension of the uplink training phase is expressed as $\tau_p < \tau_c$. We assume that the MSs transmit their assigned pilot sequence and that the channel estimation is performed locally at the APs and at the BSs in the system. The pilot sequences transmitted by the users, say $\bm{\phi}_1, \ldots, \boldsymbol{\phi}_K$, are chosen in a set of $\tau_p$ orthogonal sequences $\mathcal{P}_{\tau_p}=\left\lbrace \boldsymbol{\psi}_1, \boldsymbol{\psi}_2, \ldots, \boldsymbol{\psi}_{\tau_p} \right\rbrace$, where $\boldsymbol{\psi}_i$ is the $i$-th $\tau_p$-dimensional column sequence and $\|\boldsymbol{\psi}_i\|^2=1$, $ \forall \, i=1,\ldots, \tau_p$. In Section \ref{System_model}.\ref{Pilot_Assignment_Algorithm} we discuss the pilot assignment algorithm used in this paper.

The $m$-th AP,  during the training phase, receives the signal $\mathbf{Y}_m^{(a)} \in \mathbb{C}^{N_{\rm AP}  \times \tau_p}$, 
\begin{equation}
\mathbf{Y}_{m}^{(a)} = \ds \sum_{i =1}^K \ds \sqrt{\eta_i} \mathbf{g}_{i,m}\boldsymbol{\phi}_i^H + \mathbf{W}_{m}^{(a)} \; ,
\label{eq:y_m_a}
\end{equation}
where $\eta_i=\tau_p \overline{\eta}_i$ is the power transmitted by the $i$-th  user during the uplink training phase, and $\mathbf{W}_{m}^{(a)} \in \mathbb{C}^{N_{\rm AP}  \times \tau_p}$ contains the thermal noise contribution and out-of-cell interference at the $\ell$-th BS, with i.i.d. ${\cal CN}(0, \sigma^2_w)$ RVs as entries.
{Assuming knowledge of the large-scale fading coefficients $\beta_{k,m}$, of the vectors $\mathbf{a}_{AP}\left(\theta_{k,m}\right)$, and of the Rician factors $ K_{k,m}^{(a)}\, \forall \; m,k$, the estimates of the of the channel ${\mathbf{g}}_{k,m}$ by using the LMMSE criterion \cite{kay1993fundamentals} can be obtained as follows: 
\begin{equation}
	\widehat{\mathbf{g}}_{k,m}= \mathbf{D}_{k,m}^{(a)}\widehat{\mathbf{y}}_{k,m}^{(a)} \; ,
	\label{LMMSE_est2_a}
\end{equation}
with $\widehat{\mathbf{y}}_{k,m}^{(a)}=\mathbf{Y}_m^{(a)} \boldsymbol{\phi}_k$ and 
\begin{equation*}
\begin{array}{lllll}
\mathbf{D}_{k,m}^{(a)}= \sqrt{\eta_k} \mathbf{G}_{k,m} \mathbf{B}_{k,m}^{-1} \in \mathbb{C}^{N_{\rm AP} \times N_{\rm AP}} \, , \\
\mathbf{G}_{k,m} = \frac{\beta_{k,m}}{K_{k,m}^{(a)}+1} \left[ K_{k,m}^{(a)} \mathbf{a}_{AP}\left(\theta_{k,m}\right)\mathbf{a}_{AP}^H\left(\theta_{k,m}\right) +\mathbf{I}_{N_{\rm AP}}\right] \, , \\
\mathbf{B}_{k,m} =\sum_{i \in \mathcal{K}} {\eta_i \mathbf{G}_{i,m}\left|\boldsymbol{\phi}_i^H \boldsymbol{\phi}_k\right|^2 } + \sigma^2_w \mathbf{I}_{N_{\rm AP}}.
\end{array}
\end{equation*}
}

Similar considerations hold for the channel estimation phase at the BSs. 
The signal received at the $\ell$-th BS during the training phase $\mathbf{Y}_{\ell}^{(b)} \in \mathbb{C}^{N_{\rm BS}  \times \tau_p}$ can be indeed expressed as
\begin{equation}
\mathbf{Y}_{\ell}^{(b)} = \ds \sum_{i =1}^K \ds \sqrt{\eta_i} \mathbf{h}_{i,\ell}\boldsymbol{\phi}_i^H + \mathbf{W}_{\ell}^{(b)} \; ,
\label{eq:y_m_b}
\end{equation}
where $\mathbf{W}_{\ell}^{(b)} \in \mathbb{C}^{N_{\rm BS}  \times \tau_p}$ contains the thermal noise contribution and out-of-cell interference at the $\ell$-th BS, with i.i.d. ${\cal CN}(0, \sigma^2_w)$ RVs as entries.
From the observable $\mathbf{Y}_{\ell}^{(b)}$, the $\ell$-th BS can estimate the channel vectors $\left\{\mathbf{h}_{k,\ell}\right\}_{k=1}^K$ based on the statistics
$\widehat{\mathbf{y}}_{k,\ell}^{(b)}=\mathbf{Y}_{\ell}^{(b)} \boldsymbol{\phi}_k$ through the LMMSE approach. 
{Assuming again knowledge of the large-scale fading coefficients $\rho_{k,\ell}$, of the vectors $\mathbf{a}_{BS}\left(\phi_{k,\ell}\right)$, and of the Rician factors $ K_{k,\ell}^{(b)}\, \forall \; \ell,k$, the estimates of the channel ${\mathbf{h}}_{k,\ell}$ by using the LMMSE criterion \cite{kay1993fundamentals} can be obtained as follows: 
\begin{equation}
	\widehat{\mathbf{h}}_{k,\ell}= \mathbf{D}_{k,\ell}^{(b)}\widehat{\mathbf{y}}_{k,\ell}^{(b)} \; ,
	\label{LMMSE_est2_b}
\end{equation}
with
\begin{equation*}
\begin{array}{lllll}
\mathbf{D}_{k,\ell}^{(b)}= \sqrt{\eta_k} \mathbf{H}_{k,\ell} \mathbf{C}_{k,\ell}^{-1} \in \mathbb{C}^{N_{\rm BS} \times N_{\rm BS}} \, , \\
\mathbf{H}_{k,\ell} = \frac{\rho_{k,\ell}}{K_{k,\ell}^{(b)}+1} \left[ K_{k,\ell}^{(b)} \mathbf{a}_{BS}\left(\phi_{k,\ell}\right)\mathbf{a}_{BS}^H\left(\phi_{k,\ell}\right) +\mathbf{I}_{N_{\rm BS}}\right] \, , \\
\mathbf{C}_{k,\ell} =\sum_{i \in \mathcal{K}} {\eta_i \mathbf{H}_{i,\ell}\left|\boldsymbol{\phi}_i^H \boldsymbol{\phi}_k\right|^2 } + \sigma^2_w \mathbf{I}_{N_{\rm BS}}.
\end{array}
\end{equation*}
}

\subsection{Pilot Assignment Algorithm} \label{Pilot_Assignment_Algorithm}

We now discuss how the $\tau_p$ orthogonal pilot sequences in $\mathcal{P}_{\tau_p}$ are to be shared among the $K$ users, with $K \gg \tau_p$. In general, the distance between users using the same pilot sequence should be as large as possible to reduce the impact of \textit{pilot contamination} on the system performance. It is thus assumed that the users' positions are known to the network operator, or they can be estimated. The proposed procedure uses the well-known $k$-means clustering method \cite{K_means_steinley2006}, i.e., an iterative algorithm that is able to partition users into disjoint clusters. 
Defining the \textit{centroid} of each cluster as the mean of the positions of the APs in the cluster, 
the algorithm, accepting as input the users positions and the number of pilot sequences, is summarized in Algorithm \ref{Pilot_Assignment_Alg} and operates as follows:

\begin{itemize}
	
	\item[a.] $\lceil K/\tau_p\rceil$  centroids are chosen so that they approximately form a regular grid over the considered area, {i.e., the so-called parameter ``$k$'' in the $k$-means clustering is $\lceil K/\tau_p\rceil$.}
	
	\item[b.] Assign  each user to its nearest centroid.
	
	\item[c.] Update the centroid positions by averaging over the positions of the users belonging to each cluster.
	
	\item[d.] Repeat steps [b.] and [c.] until the positions of the centroids converge. 
	
	\item[e.] Once the users' clusters have been defined, pilot sequences are to be assigned to the users according to the following strategy.  We characterize the users in each cluster with their position relative to the centroid of the cluster to which they belong. Then, we assign the first pilot sequence to the users in each cluster that has the largest latitude (i.e. the most northern one); the second pilot sequence to the user in each cluster with the second largest latitude, and so on. The assignment procedure stops when all the the users in the system have been assigned a pilot sequence.
\end{itemize}

Regarding step [e.], its aim is to ensure that users that are assigned the same pilot sequence are not too close. The methodology that we propose here, based only on the latitude of the users and not on their full 2D coordinates, is clearly suboptimal, but has been tested to represent a good trade-off between complexity and performance.

\begin{algorithm}
	
	\caption{Pilot Assignment algorithm}
	
	\begin{algorithmic}[1]
		
		\label{Pilot_Assignment_Alg}
		\STATE  Allocate  $\lceil K/ \tau_p \rceil$ {centroids}
		so as to form an approximately regular grid over the considered area.
		\REPEAT 
		
		\STATE Assign each AP to the nearest centroid.
		
		\STATE Compute the new positions of the centroids averaging over the positions of the users belonging to the same cluster.
		
		\UNTIL convergence of the positions of the centroids or maximum number of iterations reached. 
		
		\STATE Assign the first pilot sequence to the user in each cluster that has the largest latitude (i.e. the most northern one); assign the second pilot sequence to the user in each cluster with the second largest latitude. Continue until all the the users in the system have been assigned a pilot sequence.
		
	\end{algorithmic}
	
\end{algorithm}

\subsection{The Communication Process: Downlink Data Transmission} \label{Downlink_data_transmission_section}
We assume that each user may be served by a certain subset of APs and BSs, and denote by $a_{k,m}$ and $b_{k,\ell}$ the binary variates encoding this information. In particular, $a_{k,m}=1$ ($b_{k,\ell}=1$) if the $k$-th user is served by the $m$-th AP ($\ell$-th BS), otherwise it is zero. We also assume that the coefficients
$a_{k,m}$ ($b_{k,\ell}$) are arranged in the matrix $\bA$ ($\bB$) of size $K \times M$ ($K \times L$).
The APs and the BSs treat the channel estimates as true channels and perform beamforming on the downlink. We denote by $\mathbf{w}_{k,m}^{(a)}$  ($\mathbf{w}_{k,\ell}^{(b)}$) the unit norm beamforming vector of dimension $N_{\rm AP}$ ($N_{\rm BS}$) used at the $m$-th AP ($\ell$-th BS) for the downlink transmission to the $k$-th user, and by $\eta_{k,m}^{(a)}$ ($\eta_{k,\ell}^{(b)}$) the positive scalar coefficient controlling the power transmitted by the $m$-th AP ($\ell$-th BS) to the $k$-th user. Moreover, letting $P_{m,{\rm max}}^{(a)}$ $\left(P_{\ell,{\rm max}}^{(b)}\right)$ denote the maximum transmitted power by the $m$-th AP ($\ell$-th BS), the normalized transmit powers must satisfy the constraints
\begin{equation}
\ds \sum_{k=1}^K {\eta_{k,m}^{(a)}} \leq P_{m,{\rm max}}^{(a)} \, , \quad \mbox{and} \quad
\ds \sum_{k=1}^K {\eta_{k,\ell}^{(b)}} \leq P_{\ell,{\rm max}}^{(b)} \; .
\end{equation}

{Based on the above assumptions, in downlink cell-free user-centric systems, each user receives phase-aligned contributions from the serving APs and BSs. In particular, in a generic symbol interval, the $k$-th user receives the signal expressed in Eq. \eqref{eq:received_data_MS_UC_co-existence} at the top of the next page, with $x_k^{\rm DL}$ the unit power data symbols intended for the $k$-th user, and $z_k$ the additive white Gaussian noise (AWGN) with entries ${\cal CN}(0, \sigma^2_z)$. 
\begin{figure*}[t]
{
\begin{equation}
\begin{array}{llll}
r_k^{\rm DL} & = \left( \ds \sum_{m=1}^M a_{k,m} \ds \sqrt{\eta_{k,m}^{(a)}} \mathbf{g}_{k,m}^H  \mathbf{w}_{k,m}^{(a)} + \ds \sum_{\ell=1}^L b_{k,\ell} \ds \sqrt{\eta_{k,\ell}^{(b)}}  \mathbf{h}_{k,\ell}^H  \mathbf{w}_{k,\ell}^{(b)}\right) x_k^{\rm DL} \\ & \quad + 
\ds \sum_{\substack{j=1 \\ j\neq k}}^K \left( \ds \sum_{m=1}^M a_{j,m} \ds \sqrt{\eta_{j,m}^{(a)}}  \mathbf{g}_{k,m}^H  \mathbf{w}_{j,m}^{(a)} + \ds \sum_{\ell=1}^L b_{j,\ell} \ds \sqrt{\eta_{j,\ell}^{(b)}} \mathbf{h}_{k,\ell}^H  \mathbf{w}_{j,\ell}^{(b)}\right) x_j^{\rm DL} +  
z_k\; ,
 \end{array}
\label{eq:received_data_MS_UC_co-existence}
\end{equation}
\hrulefill}
\end{figure*}
Given \eqref{eq:received_data_MS_UC_co-existence}, the $k$-th user downlink SINR, $\gamma_k^{\rm DL}$ say,  conditioned on the fast fading channel realizations and on the channel vector estimates can be easily shown to be written as
\begin{equation}
\!\!	\gamma_k^{\rm DL}\!=\!\frac{\ds \left|\ds \sum_{m=1}^M \!\! a_{k,m} \ds \sqrt{\eta_{k,m}^{(a)}} \mathbf{g}_{k,m}^H  \mathbf{w}_{k,m}^{(a)} \!+\! \ds \sum_{\ell=1}^L b_{k,\ell} \ds \sqrt{\eta_{k,\ell}^{(b)}} \mathbf{h}_{k,\ell}^H  \mathbf{w}_{k,\ell}^{(b)}\right|^2}{\ds \sum_{\substack{j=1 \\ j\neq k}}^K \!\left| \ds \! \sum_{m=1}^M \! \! a_{j,m} \ds \!\sqrt{\!\eta_{j,m}^{(a)}} \mathbf{g}_{k,m}^H  \!\mathbf{w}_{j,m}^{(a)} \! \!+ \!\!\ds \sum_{\ell=1}^L \! b_{j,\ell}\! \ds \sqrt{\!\eta_{j,\ell}^{(b)}} \mathbf{h}_{j,\ell}^H  \!\mathbf{w}_{j,\ell}^{(b)}\right|^2\!\!+\!\sigma^2_z} \, .
\label{eq:gamma_k}\end{equation}}

\section{Downlink spectral efficiency bounds} \label{DL_SE_Bounds}
The performance measure that is considered in this work is the per-user downlink rate, which can be obtained by multiplying the downlink per-user spectral efficiency (SE) by the signal bandwidth. Unfortunately, the exact closed-form expression of the downlink SE is not available, and suitable bounds are to be adopted. Assuming a Gaussian approximation for the interference,  Gaussian distributed very long codewords, and perfect channel state information (PCSI), the ergodic SE for the $k$-th user is written as
\begin{equation}
\text{SE}_{k}^{\rm DL}=\mathbb{E}\left[\log_2 \left( 1+
\gamma_k^{\rm DL}
 \right) \right]\, ,
\label{SE_DL_PCSI_co-existence}
\end{equation}
where the statistical expectation is to be performed with respect to the channel coefficients. 

In the practical case of imperfect channel state information (ICSI), the channel coefficients are estimated in the uplink, and the estimated values are used to compute the beamformers. Let us thus denote by  $\widehat{\gamma}_k^{\rm DL}$ the $k$-th user SINR expression in \eqref{eq:gamma_k}, where the beamformers $\mathbf{w}_{k,m}^{(a)}$ and $\mathbf{w}_{k,\ell}^{(b)}$, for all values of $k, m$ and $\ell$, are computed based on the available channel estimates. It can be then shown that, for the ICSI case, an upper bound to the achievable SE for the user $k$ is written as
\begin{equation}
\text{SE}_{k, {\rm UB}}^{\rm DL}=\ds\frac{{\tau}_{\rm d}}{\tau_c} \mathbb{E}\left[\log_2 \left( 1+  \widehat{\gamma}_k^{\rm DL} \right)\right] \, ,
\label{SE_DL_ICSI_co-existence_UpperBound}
\end{equation}
where ${\tau}_{\rm d}$ is the length (in time-frequency samples) of  the downlink data transmission phase. Moreover, in \eqref{SE_DL_ICSI_co-existence_UpperBound}, 
$\widehat{\gamma}_k^{\rm DL}$ has the same expression as ${\gamma}_k^{\rm DL}$ in \eqref{eq:gamma_k}, with the only difference that the beamformers are computed using the estimated channel values and not the true ones.

In addition to the upper bound \eqref{SE_DL_ICSI_co-existence_UpperBound}, there are some lower bound expressions for the per user downlink SE. These include the Use-and-then-Forget (UatF) bound \cite{marzetta2016fundamentals}, and the Caire bound in \cite{caire2018ergodic}. The former bound had been originally conceived for massive MIMO systems, and was extended to cell-free massive MIMO systems in \cite{Ngo_CellFree2017}. However, this bound has been shown to be somehow loose for cell-free massive MIMO systems, since it works well in the presence of channel hardening, a phenomenon that takes place in the massive MIMO scenario where BSs are equipped with a large number of co-located antennas. Moreover, a closed-form expression of the beamformer is available only for a limited number of beamformer types. To avoid these drawbacks, in \cite{caire2018ergodic} two alternative lower bounds were proposed; these, however, are tight when the channel coherence block length is large with respect to the number of users, and in the case where interference is nearly entirely eliminated by zero-forcing beamforming. Clearly, these assumptions are hardly verified in the practically relevant case of an overloaded network with a large number of active users and a non-negligible level of interference. Consequently, in this paper, we will use the upper boud \eqref{SE_DL_ICSI_co-existence_UpperBound} as our key performance indicator. In fact, the bound tends to become tight when the channel estimation error decreases; moreover, although showing SE values not practically achievable, it is still useful when comparing different network deployments, network cooperation levels, and different implementations of beamformers and power control rules.

\section{System design} \label{System_design}
In this section, we discuss the UE-AP/BS association rules considered, the adopted beamformers, and power control strategies.

\subsection{Basic UE-AP/BS association rules} \label{Association_rules}
The optimal association between each UE and the serving APs and BSs descends from the adopted optimization criterion and is generally a NP-hard combinatorial problem. Accordingly, heuristic association rules are considered in practice based on the large-scale fading coefficients. 

For the \textit{MC-mMIMO} scenario (i.e., multicell massive MIMO without BS cooperation), each UE is simply associated to the BS with the highest value of the large-scale fading coefficient. Otherwise stated, letting $\ell^*_k=\arg \max_{\ell} \rho_{k,\ell}$, we have $b_{k,\ell^*_k}=1$ and $b_{k,\ell}=0$, for all $\ell \neq \ell^*_k$, and the matrix $\bA$ consists of all zeros. 

For the \textit{HORIZONTAL-Coop} scenario, each UE can connect to a certain number, say $\widetilde{N}_{\rm BS}$, of BSs  or to a certain number, say $\widetilde N_{\rm AP}$, of APs. The association rule has to take into account the strength of the large-scale fading coefficients and the number of antennas at the APs and BSs. A suitable association rule can be obtained as follows:
\begin{enumerate}
	\item Let $O_k^\beta \, : \, \{1,\ldots, M \} \rightarrow \{1,\ldots, M \}$ denote the sorting operator for the vector $\left[\beta_{k,1},\ldots, \beta_{k,M}\right]$, such that $\beta_{k,O^\beta_k(1)} \geq \beta_{k,O^\beta_k(2)} \geq \ldots \geq \beta_{k,O^\beta_k(M)}$. Similarly, let $O_k^\rho \, : \, \{1,\ldots, L \} \rightarrow \{1,\ldots, L \}$ denote the sorting operator for the vector $\left[\rho_{k,1},\ldots, \rho_{k,L}\right]$, such that $\rho_{k,O^\rho_k(1)} \geq \rho_{k,O^\rho_k(2)} \geq \ldots \geq \rho_{k,O^\rho_k(L)}$.
	\item If the following condition holds:
	\begin{equation}
		\ds N_{\rm AP} \sum_{i=1}^{\widetilde N_{\rm AP}} \beta_{k,O^\beta_k(i)} \geq N_{\rm BS}
		 \sum_{i=1}^{\widetilde N_{\rm BS}} \rho_{k,O^\rho_k(i)}\; ,
		 \label{eq:condizioneAP-UE}
	\end{equation}
    then the $k$-th UE is to be associated to the APs with index $O^\beta_k(1), O^\beta_k(2), \ldots, O^\beta_k(\widetilde N_{\rm AP})$. Consequently, the coefficients $a_{k,O^\beta_k(1)}, \ldots, a_{k,O^\beta_k(\widetilde N_{\rm AP})}$ should be set equal to 1, and all other entries in the matrix $\bA$ and the entire matrix $\bB$ should be set to zero. 
    \item If, on the contrary, \eqref{eq:condizioneAP-UE} does not hold, then the $k$-th UE is to be associated to the BSs with index $O^\rho_k(1), O^\rho_k(2), \ldots, O^\rho_k(\widetilde N_{\rm BS})$. Consequently, the coefficients $b_{k,O^\rho_k(1)}, \ldots, b_{k,O^\rho_k(\widetilde N_{\rm BS})}$ should be set equal to 1, and all other entries in the matrix $\bB$ and the entire matrix $\bA$ should be set to zero. 
\end{enumerate}

For the \textit{HET-noCoop} scenario, where each UE can connect to either one BS or to one AP, the association rule UE-AP/BS is a special case of the one outlined for the above \textit{HORIZONTAL-Coop} scenario, obtained by setting
$\widetilde N_{\rm BS}=\widetilde N_{\rm AP}=1$.

Finally, in the \textit{FULL-Coop} scenario, each UE is cooperatively served by $\widetilde N_{\rm AP}$ APs and by $\widetilde N_{\rm BS}$ BSs. Also for this scenario the large-scale fading coefficients are considered for selecting the serving APs and BSs. In particular, using again the sorting operators $O_k^\beta$ and $O_k^\rho$, we have that the association variables $a_{k,O^\beta_k(1)}, \ldots, a_{k,O^\beta_k(\widetilde N_{\rm AP})}$ and $b_{k,O^\rho_k(1)}, \ldots, b_{k,O^\rho_k(\widetilde N_{\rm BS})}$ are set equal to one, while the remaining entries of the matrices $\bA$ and $\bB$ are set to zero{\footnote{{The cooperation between APs and/or BSs can be affected by timing synchronization errors whose consideration is beyond the scope of this paper. However, it is worth noting that, if every AP or BS is synchronized with its neighbors and the UE synchronizes with its Master AP or Master BS in the initial access, the model used in the paper can be assumed reasonable for analysis\cite{bjornson2020scalable}. }}}.

The above described association rules are, as already highlighted, simple and heuristic. Thus, alternative and more sophisticated rules can be conceived. In particular, the association could take into account not only the channel strength, as the described policies do, but also the interference level. Moreover, a load balancing constraint could also be introduced, in order to prevent the case in which a single AP or BS is connected to a too large number of devices. The final part of this paper will provide results along these lines.

\subsection{Downlink beamforming} \label{Local_precoding_Section}
Once the UE-AP/BS association rule has been implemented, the set
$\mathcal{K}_m$ ($\mathcal{J}_{\ell}$) can be defined for the generic $m$-th AP ($\ell$-th BS). In particular, $\mathcal{K}_m$ is the set of users served by the $m$ -th AP, that is, $ \mathcal{K}_m =\lbrace k \, : \, a_{k,m}=1 \rbrace$. Similarly,   $\mathcal{J}_{\ell}$ is the set of users served by the $\ell$-th BS, that is, $ \mathcal{J}_\ell =\lbrace k \, : \, b_{k,\ell}=1 \rbrace$.

In this paper, the following three downlink beamformers will be considered. 

\textit{Maximum ratio transmission (MRT)}: In this case, the $m$-th AP ($\ell$-th BS),  simply uses the local channel estimate to send data to the $k$-th user, i.e. we have
	\begin{equation}
	\ds \mathbf{w}_{k,m}^{(a), {\rm MRT}} =\ds \frac{\widehat{\mathbf{g}}_{k,m}^H }{\norm{\widehat{\mathbf{g}}_{k,m}}} \, , \qquad \ds \mathbf{w}_{k,\ell}^{(b), {\rm MRT}} =\ds \frac{\widehat{\mathbf{h}}_{k,\ell}^H }{\norm{\widehat{\mathbf{h}}_{k,\ell}}} \, ,
	\label{eq:MRT_beamformer}
	\end{equation}
where the assumption that $k \in \mathcal{K}_m$ and that $k \in \mathcal{J}_{\ell}$ is implicitly done. 

\textit{Partial zero-forcing (PZF)}: In this case, we assume that each AP performs a local PZF downlink precoding, i.e. it nulls its interference contribution towards some selected users. Consider the $m$-th AP and assume that $k \in \mathcal{K}_m$. Denote by $N_{\rm PZF}^{\rm AP}$ the number of users to be protected by the interference generated by each AP\footnote{Recall that $N_{\rm PZF}^{\rm AP}$ must not exceed the number of antennas available at the AP.} when transmitting to UE $k$. Denote by $j_{k,m}(1), \ldots, j_{k,m}(N_{\rm PZF}^{\rm AP})$ the index of the users to be protected\footnote{A customary choice may be to select the users, other than the $k$-th, with the strongest large-scale fading coefficient towards the $m$-th AP.}; the PZF precoder to be used at $m$-th AP to transmit to the $k$-th user is computed as follows:
\begin{enumerate}
	\item The matrix $\mathbf{I}_{k,m}=[\widehat{\mathbf{g}}_{j_{k,m}(1),m}, \ldots, 
	\widehat{\mathbf{g}}_{j_{k,m}(N_{\rm PZF}^{\rm AP}),m}]$, containing the estimated channel of the users to be protected is formed.
	\item Let $\widetilde{\mathbf{I}}_{k,m}= \mbox{orth}\left[\mathbf{I}_{k,m}\right]$ be a matrix that contains in its columns an orthonormal basis for the column span of $\mathbf{I}_{k,m}$.
	\item Define the PZF 
	$$ \mathbf{w}_{k,m}^{(a), {\rm PZF}}=\left( \mathbf{I}_{N_{\rm AP}} - 
	\widetilde{\mathbf{I}}_{k,m} \widetilde{\mathbf{I}}_{k,m}^H \right)\widehat{\mathbf{g}}_{k,m}$$
	\item Carry out the beamformer normalization:\\ $\mathbf{w}_{k,m}^{(a), {\rm PZF}} \leftarrow \mathbf{w}_{k,m}^{(a), {\rm PZF}}/\|\mathbf{w}_{k,m}^{(a), {\rm PZF}}\|$.	
\end{enumerate} 
A similar procedure is used to determine the beamformer in $\ell$-th BS for transmission to UE $k$. 
Denote by $N_{\rm PZF}^{\rm BS}<N_{\rm BS}$ the number of users to be protected by the interference generated by BS $\ell$ when transmitting to UE $k$. Denote by $j_{k,\ell}(1), \ldots, j_{k,\ell}(N_{\rm PZF}^{\rm BS})$ the index of the users to be protected\footnote{Again, these can be  the users, other than the $k$-th, with the strongest large-scale fading coefficient towards the $\ell$-th BS.}; the PZF precoder to be used at the $\ell$-th BS to transmit to $k$-th user thus requires the formation of the interference matrix  $\mathbf{I}_{k,\ell}=[\widehat{\mathbf{h}}_{j_{k,\ell}(1),\ell}, \ldots, 
	\widehat{\mathbf{h}}_{j_{k,\ell}(N_{\rm PZF}^{\rm BS}),\ell}]$, containing the estimated channel of the users to be protected, the computation of an orthonormal basis $\widetilde{\mathbf{I}}_{k,\ell}= \mbox{orth}\left[\mathbf{I}_{k,\ell}\right]$, and of the beamformer: 
Define the PZF 
	$$ \mathbf{w}_{k,\ell}^{(b), {\rm PZF}}=\left( \mathbf{I}_{N_{\rm BS}} - 
	\widetilde{\mathbf{I}}_{k,\ell} \widetilde{\mathbf{I}}_{k,\ell}^H \right)\widehat{\mathbf{h}}_{k,\ell}$$
	Finally, the obtained beamformer must be made of unit-energy: \\ $\mathbf{w}_{k,\ell}^{(b), {\rm PZF}} \leftarrow \mathbf{w}_{k,\ell}^{(b), {\rm PZF}}/\|\mathbf{w}_{k,\ell}^{(b), {\rm PZF}}\|$

\textit{Minimum mean squares error (MMSE)}: The MMSE beamforming vector is obtained exploiting the duality principle in massive MIMO systems and minimizing the conditional mean-squares error (MSE) in the uplink decoding\cite{Bjornson_Sanguinetti_book}. Assuming the LMMSE channel estimation discussed in Section \ref{System_model}.\ref{LMMSE_channel_estimation},  the MMSE beamforming vectors computed at the $m$-th AP and at the $\ell$-th BS for the $\mathcal{K}_m(i)$-th user and the $\mathcal{J}_{\ell}(i)$-th user are reported at the top of next page in Eqs. \eqref{w_a_MMSE} and \eqref{w_b_MMSE}, respectively.
	
	\begin{figure*}		
			\begin{equation}
			\ds \mathbf{w}_{\mathcal{K}_m(i),m}^{(a), {\rm MMSE}} =\frac{\left \lbrace\ds  \sum_{j \in \mathcal{K}_m} \eta_j \left[\widehat{\mathbf{g}}_{j ,m} \widehat{\mathbf{g}}_{j ,m}^H + \mathbf{G}_{j,m} - \widehat{\mathbf{G}}_{j,m} \right]+ \sigma^2_z \mathbf{I}_{N_{\rm AP}}  \right \rbrace^{-1} \widehat{\mathbf{g}}_{\mathcal{K}_m(i) ,m}}{ \norm{ \left \lbrace\ds  \sum_{j \in \mathcal{K}_m} \eta_j \left[\widehat{\mathbf{g}}_{j ,m} \widehat{\mathbf{g}}_{j ,m}^H + \mathbf{G}_{j,m} - \widehat{\mathbf{G}}_{j,m}  \right]+ \sigma^2_z \mathbf{I}_{N_{\rm AP}}  \right \rbrace^{-1} \widehat{\mathbf{g}}_{\mathcal{K}_m(i) ,m}} } \, ,
\label{w_a_MMSE}		
  \end{equation}
		
			\begin{equation}
			\ds \mathbf{w}_{\mathcal{J}_{\ell}(i),\ell}^{(b), {\rm MMSE}} =\frac{\left \lbrace\ds  \sum_{j \in \mathcal{J}_{\ell}} \eta_j \left[\widehat{\mathbf{h}}_{j ,\ell} \widehat{\mathbf{h}}_{j ,\ell}^H + \mathbf{H}_{j,\ell} - \widehat{\mathbf{H}}_{j,\ell} \right]+ \sigma^2_z \mathbf{I}_{N_{\rm BS}}  \right \rbrace^{-1} \widehat{\mathbf{h}}_{\mathcal{J}_{\ell}(i) ,\ell}}{ \norm{ \left \lbrace\ds  \sum_{j \in \mathcal{J}_{\ell}} \eta_j \left[\widehat{\mathbf{h}}_{j ,\ell} \widehat{\mathbf{h}}_{j ,\ell}^H + \mathbf{H}_{j,\ell} - \widehat{\mathbf{H}}_{j,\ell} \right]+ \sigma^2_z \mathbf{I}_{N_{\rm BS}}  \right \rbrace^{-1} \widehat{\mathbf{h}}_{\mathcal{J}_{\ell}(i) ,\ell}} } \, ,
   \label{w_b_MMSE}	
		\end{equation}
		\hrule
		\hfill
	\end{figure*}

\subsection{Power allocation strategies}
In order to keep the system complexity at a reasonable level and to come up with a practically deployable system implementation, a simple fractional power allocation (FPA) strategy is considered here. 
We denote as $P_{m,{\rm max}}^{(a)}$ and $P_{\ell,{\rm max}}^{(b)}$ the maximum powers available at the $m$-th AP and at the $\ell$-th BS for the downlink transmission, respectively.
The generic $m$-th AP ($\ell$-th BS) serves the UEs in $\mathcal{K}_m$ ($\mathcal{J}_{\ell}$), i.e. those UEs such that $a_{k,m}=1$ ($b_{k,\ell}=1$), with the following power
	\begin{equation}
	\eta_{k,m}^{(a)}=\frac{P_{m,{\rm max}}^{(a)}/\beta_{k,m}^{\alpha}}{\ds \sum_{j=1}^K \frac{a_{j,m}}{\beta_{j,m}^{\alpha}}}, \quad
	\eta_{k,\ell}^{(b)}=\frac{P_{\ell,{\rm max}}^{(b)}/\rho_{k,\ell}^{\alpha}}{\ds \sum_{j=1}^K \frac{b_{j,\ell}}{\rho_{j,\ell}^{\alpha}}}
	\label{FPA_powers}
	\end{equation}
In \eqref{FPA_powers}, $\alpha$ is a real parameter, which usually takes a value in the range $[-1,1]$. Stategies \eqref{FPA_powers} define thus a family of power control strategies, depending on the chosen value of $\alpha$. Notice that for $\alpha=0$ we have the simple uniform power allocation, where each AP/BS equally divides the available power among the UEs he is serving. Positive values of $\alpha$ usually favor fairness between users (that is, more power is given to UEs with weaker channels), while negative values of $\alpha$ are beneficial for UEs with the best channel conditions.

\section{FULL-Coop scenario: joint precoding scheme} \label{Joint_PZF_Section}
We now consider, for the FULL-Coop scenario, a joint precoding scheme exploiting the cooperation between BSs and APs. 
In fact, in practical implementations, the centralized computation of the beamformer must take into account the local power budget constraints at each AP and each BS. In the following, we provide a procedure for fulfilling this constraint. 
For the sake of brevity, we now discuss only the joint partial zero-forcing (JPZF) but all the precoding schemes discussed above for the local processing can be considered.
For computing the centralized beamformer, one scaling factor must be used for each UE in the network. We thus assume that:
\begin{equation}
	\eta_{k,m}^{(a)}=\eta_{k,\ell}^{(b)}=\eta_k \; \forall \, m \in \mathcal{M}_k, \, \ell \in \mathcal{B}_k,
\end{equation}
$k=1,\ldots,K$, where $\eta_k$ is now the parameter to be determined, and the sets $\mathcal{M}_k$ and $\mathcal{B}_k$ contain the $\widetilde N_{\rm AP}$ APs and the $\widetilde N_{\rm BS}$ BSs serving the $k$-th user, respectively. 
In order to design the precoding vector for the $k$-th UE, we define for any $k$ the following vector:
$$
\widetilde{\mathbf{u}}_j^{(k)}= \left[  \widehat{\mathbf{g}}_{j ,\mathcal{M}_k(1)}^T, \ldots, \widehat{\mathbf{g}}_{j ,\mathcal{M}_k(\widetilde N_{\rm AP})}^T,  \widehat{\mathbf{h}}_{j ,\mathcal{B}_k(1)}^T, \ldots, \widehat{\mathbf{h}}_{j ,\mathcal{B}_k(\widetilde N_{\rm BS})}^T\right]^T\!\!\!\!,$$
$ \forall k,j=1,\ldots, K.$
Next, we denote by $\mathcal{R}_{k}$ the set containing the $R_{\rm JPZF}$ vectors with the highest norm in the set $ \left \lbrace  \widetilde{\mathbf{u}}_j^{(k)} \right \rbrace_{j \neq k} $, where $R_{\rm JPZF}$ represents the number of users to be protected from the interference generated by the downlink transmission to the $k$-th user.
The $\left( N_{\rm AP} \widetilde N_{\rm AP} + N_{\rm BS} \widetilde N_{\rm BS}\right) $-dimensional beamforming vector $\mathbf{w}_{k}^{\rm JPZF}$ for the $k$-th UE has to fulfil the following conditions:
\begin{equation}
	\left\{ 
	\begin{array}{lll} 
		\widetilde{\mathbf{u}}_k^{(k)} \mathbf{w}_{k}^{\rm JPZF}>0 \;, \\
		\widetilde{\mathbf{u}}_j^{(k)} \mathbf{w}_{k}^{\rm JPZF} =0 \;, \; \forall \; \; j \in  \mathcal{R}_{k}\; ,
	\end{array} 
	\right.
	\label{eq:downlink_precoder_JPZF}
\end{equation}
which require that that $N_{\rm AP} \widetilde N_{\rm AP} + N_{\rm BS} \widetilde N_{\rm BS} \geq R_{\rm JPZF}$.

To effectively determine $\mathbf{w}_{k}^{\rm JPZF}$, we first form the matrix of dimension $\left[ \left(N_{\rm AP} \widetilde N_{\rm AP} + N_{\rm BS} \widetilde N_{\rm BS} \right) \times R_{\rm JPZF} \right]$ $\mathbf{I}_{k}^{\rm JPZF}$, whose columns are the vectors $
\left \lbrace  \widetilde{\mathbf{u}}_j^{(k)} \right \rbrace_{ j \in \mathcal{R}_{k}}$. Then, the $\left(N_{\rm AP} \widetilde N_{\rm AP} + N_{\rm BS} \widetilde N_{\rm BS} \right) $-dimensional beamforming vector $\mathbf{w}_{k}^{\rm JPZF}$ can be obtained as 

\begin{equation}
	\mathbf{w}_{k}^{\rm JPZF}= \frac{\left[ \mathbf{I}_{N_{\rm AP} \widetilde N_{\rm AP} + N_{\rm BS} \widetilde N_{\rm BS}} - \widetilde{\mathbf{I}}_{k}^{\rm JPZF} \widetilde{\mathbf{I}}_{k}^{{\rm JPZF},H} \right] \widetilde{\mathbf{u}}_k^{(k)}}{\norm{\left[ \mathbf{I}_{N_{\rm AP} \widetilde N_{\rm AP} + N_{\rm BS} \widetilde N_{\rm BS}} - \widetilde{\mathbf{I}}_{k}^{\rm JPZF} \widetilde{\mathbf{I}}_{k}^{{\rm JPZF},H} \right] \widetilde{\mathbf{u}}_k^{(k)}}}\,,
	\label{Centralized_PZF}
\end{equation}
where $\widetilde{\mathbf{I}}_{k}^{\rm JPZF}= \text{orth}\left( \mathbf{I}_{k}^{\rm JPZF} \right)$.

In order to allocate the same power to all the users and ensure that the maximum transmit power per AP and per BS is fulfilled, the following procedure is adopted.
We collect the beamforming vectors in the $\left[ \left( N_{\rm AP} M + N_{\rm BS} L\right)\times  K \right]$-dimensional matrix $\mathbf{Q}$ using the pseudo-code reported in Algorithm \ref{Q_definition}. As we can note from Algorithm \ref{Q_definition}, matrix $\mathbf{Q}$ contains the partition of the joint precoding vector transmitted by each AP and BS in the cooperation network.

\begin{algorithm}[!t]
	
	\caption{Definition of matrix $\mathbf{Q}$}
	
	\begin{algorithmic}[1]
		
		\label{Q_definition}
		
		\FOR {$k=1,\ldots, k$}
		
		\STATE  Initialize $\rm{ind}=0$
		
		\FOR {$m=1,\ldots, M$}
		
		\IF {$a_{k,m}=1$}
		\STATE {\small $\!\!\!\!\!\!\!\!\!\!\!\mathbf{Q}\!\left[(m-1)N_{\rm AP}\!+\!1\!:\!mN_{\rm AP},k \right]\!\!=\!\!\mathbf{w}_{k}^{\rm JPZF}\!\!\left[{\rm ind}\!+\!1\!:\!{\rm ind}\!+\! N_{\rm AP}\right]\!\!\!$}
		\STATE ${\rm ind}={\rm ind}+N_{\rm AP}$ 
		\ENDIF
		
		\ENDFOR
		
		\FOR {$\ell=1,\ldots, L$}
		
		\IF {$b_{k,\ell}=1$}
		\STATE {\small $\mathbf{Q}\left[N_{\rm AP}M+(\ell-1)N_{\rm BS}+1:N_{\rm AP}M+\ell N_{\rm BS},k\right]=\mathbf{w}_{k}^{\rm JPZF}\left[{\rm ind}+1:{\rm ind}+N_{\rm BS}\right]$}
		\STATE ${\rm ind}={\rm ind}+N_{\rm BS}$ 
		\ENDIF
		
		\ENDFOR

		\ENDFOR
	\end{algorithmic}
	
\end{algorithm}

Upon defining the matrix $\mathbf{Q}$, the equal-stream power allocation can be solved as follows. Upon defining the 
$K$-dimensional vector $\bm{\eta}=\left[ \eta_1, \ldots, \eta_{K} \right]^T$, the following power constraints must be satisfied
\begin{subequations}\label{Prob:Equal_power}
	\begin{align}
		&\mathbf{1}_{1\times N_{\rm AP}} \text{Re} \left \lbrace  \mathbf{Q}_{m, {\rm AP}} \odot \mathbf{Q}^*_{m, {\rm AP}} \right \rbrace \bm{\eta} \! \leq \! P_{m,{\rm max}}^{(a)}, \label{Prob:aEq_power}
		\quad  m=1,\ldots, M \\
		&\mathbf{1}_{1\times N_{\rm BS}} \text{Re} \left \lbrace \mathbf{Q}_{\ell, {\rm BS}} \odot \mathbf{Q}^*_{\ell, {\rm BS}} \right \rbrace 
		\bm{\eta} \leq  P_{\ell,{\rm max}}^{(b)}, \; \; \ell=1,\ldots, L \label{Prob:bEq_power} \\ 
		&\eta_{1}\norm{\mathbf{Q}(:,1)}^2=\ldots= \eta_{K}\norm{\mathbf{Q}(:,K)}^2 \label{Prob:cEq_power} 
	\end{align}
\end{subequations}
where
\begin{equation*}
	\begin{array}{ll}
	&\mathbf{Q}_{m, {\rm AP}}= \mathbf{Q}\left[(m-1)N_{\rm AP}+1:mN_{\rm AP},:\right], \; \text{and} \\
	&\mathbf{Q}_{\ell, {\rm BS}}= \mathbf{Q}\left[N_{\rm AP}M\!\!+\!\!(\ell\!-\!1)N_{\rm BS}\!\!+\!\!1:N_{\rm AP}M\!+ \!\ell N _{\rm BS},:\right].
\end{array}
\end{equation*}
The lines \eqref{Prob:aEq_power}-\eqref{Prob:bEq_power} represent the per-AP and per-BS power constraints and line \eqref{Prob:cEq_power} represents the equal stream power constraint.
From  \eqref{Prob:cEq_power}, it descends $\bm{\eta}=\eta \mathbf{1}_{K \times 1}$, and the constraints \eqref{Prob:aEq_power}-\eqref{Prob:bEq_power} can be easily satisfied by letting
\begin{equation}
	\eta= \min \left( \frac{P_{1,{\rm max}}^{(a)}}{\widetilde{q}_1^{(a)}},\ldots, \frac{P_{M,{\rm max}}^{(a)}}{\widetilde{q}_M^{(a)}}, \frac{P_{1,{\rm max}}^{(b)}}{\widetilde{q}_1^{(b)}},\ldots, \frac{P_{L,{\rm max}}^{(b)}}{\widetilde{q}_L^{(L)}} \right)\, ,
	\label{Eta_Eq_power}
\end{equation} 
where
\begin{equation}
	\widetilde{q}_m^{(a)}=\mathbf{1}_{1\times N_{\rm AP}} \text{Re} \left \lbrace  \mathbf{Q}_{m, {\rm AP}} \odot \mathbf{Q}^*_{m, {\rm AP}} \right \rbrace \mathbf{1}_{K \times 1}\, ,
	\label{q_tilde_a}
\end{equation} 
and
\begin{equation}
	\begin{array}{llll}
		\widetilde{q}_\ell^{(b)}=\mathbf{1}_{1\times N_{\rm BS}} \text{Re} \left \lbrace \mathbf{Q}_{\ell, {\rm BS}} \odot \mathbf{Q}^*_{\ell, {\rm BS}} \right \rbrace \mathbf{1}_{K \times 1}\, .
	\end{array}
	\label{q_tilde_b}
\end{equation} 

{\subsection{Computational complexity for pre-coder computation}
In the following, we evaluate the complexity of the precoder computation in terms of complex multiplications required. The calculation of $\mathbf{w}_{k}^{{\rm JPZF}}$ requires the calculation of $\widetilde{\mathbf{I}}_{k}^{\rm JPZF}= \text{orth}\left[\mathbf{I}_{k}^{\rm JPZF}\right]$, i.e., 
the matrix containing on its columns an orthonormal basis for the column span of the $\left[ \left(N_{\rm AP} \widetilde N_{\rm AP} + N_{\rm BS} \widetilde N_{\rm BS} \right) \times R_{\rm JPZF} \right]$-dimensional matrix $\mathbf{I}_{k}^{\rm JPZF}$. This operation can be shown to require a number of complex multiplications equal to $O\left[ R_{\rm JPZF}^2 \left( N_{\rm AP} \widetilde N_{\rm AP} + N_{\rm BS} \widetilde N_{\rm BS}\right)\right]$. Then, a projection of an $\left(N_{\rm AP} \widetilde N_{\rm AP} + N_{\rm BS} \widetilde N_{\rm BS} \right)$ dimensional vector onto the orthogonal complement to the subspace spanned by the columns of $\widetilde{\mathbf{I}}_{k}^{\rm JPZF}$ is to be performed. This operation requires $O\left[ R_{\rm JPZF} \left( N_{\rm AP} \widetilde N_{\rm AP} + N_{\rm BS} \widetilde N_{\rm BS}\right)\right]$ complex multiplications. The dominant term determining the complexity for the computation of the centralized precoders is thus $O\left[ R_{\rm JPZF}^2 \left( N_{\rm AP} \widetilde N_{\rm AP} + N_{\rm BS} \widetilde N_{\rm BS}\right)\right]$.}

In contrast, in the case of local processing, the precoders $\mathbf{w}_{k,\ell}^{(b), {\rm PZF}}$ are calculated in $\widetilde{N}_{\rm BS}$ BS and the precoders $\mathbf{w}_{k,m}^{(a), {\rm PZF}}$ are calculated in $\widetilde{N}_{\rm AP}$ APs, leading to the cost of complex multiplications $ \widetilde{N}_{\rm BS}O\left( \left[N_{\rm PZF}^{\rm BS}\right]^2  N_{\rm BS} \right) + \widetilde{N}_{\rm AP} O\left( \left[N_{\rm PZF}^{\rm AP}\right]^2  N_{\rm AP}^2 \right) $.

As an example, assuming $N_{\rm BS}=32$, $N_{\rm AP}=8$, $N_{\rm PZF}^{\rm BS}=N_{\rm BS}/2=16$, $N_{\rm PZF}^{\rm AP}=N_{\rm AP}/2=4$, $\widetilde{N}_{\rm BS}=3$, $\widetilde{N}_{\rm AP}=6$, $R_{\rm JPZF}= \left( N_{\rm AP} \widetilde N_{\rm AP} + N_{\rm BS} \widetilde N_{\rm BS}\right)/2=72$, the computational complexity of the centralized precoding is $O\left[ 72^2 \cdot 144\right] \approx O((7.4 \cdot 10^5))$ complex multiplications while the complexity of the local precoding is $3 O\left( 16^2 \cdot  32 \right) + 6 O\left( 4^2  \cdot 8 \right) \approx \; O(2.5 \cdot 10^4) $ complex multiplications.

\section{FULL-Coop scenario: consideration of limited-capacity fronthaul constraint}\label{Fronthaul_Section}
Considering again the FULL-Coop scenario,  we now provide a heuristic procedure aimed at reducing the fronthaul load. We consider both the local processing at the AP/BS discussed in Section \ref{System_design}.\ref{Local_precoding_Section} and the joint processing presented in Section \ref{Joint_PZF_Section}, and we also take into account the use of multiple subcarriers. We assume that the generic $n$-th AP/BS is connected to the CPU via a fronthaul link with a capacity of $\bar{F}_n, \forall n$.
First of all, we formulate the total FH requirement at the $n$-th AP/BS as the sum of the \textit{FH data} and \textit{FH weights} requirements, and we aim at designing a UE association algorithm such that the fronthaul link capacity is not exceeded, i.e.,	
		\begin{equation}
		\text{F}_n=\text{F}_n^{(D)}+\text{F}_n^{(W)} \leq \bar{F}_n
		\label{eq:FH_requirement}\end{equation}
where $\text{F}_n^{(D)}$ and $\text{F}_n^{(W)}$ are the share of the FH requirement related to data payload and beamforming weights (i.e., precoding vectors), respectively. 

Under the assumption that all users are served over \textit{all} the subcarriers, if the $k$-th UE is served by the $n$-th AP/BS, the CPU needs to send over the FH link to the $n$-th AP/BS the $k$-th UE information symbols for all the subcarriers.
Accordingly, the data rate related to the \textit{FH data} is written as
		\begin{equation}
	\text{F}_n^{(D)}= \ds \sum_{k=1}^K  \widetilde{a}_{k,n}  \log_2 (M_{\rm QAM}) N_{\rm RB} \frac{N_{\rm sc}^{(RB)}}{\tau_{\rm data}} \frac{N_{\rm sc}^{(OFDM)}}{\eta_{\rm CPRI}} \; ,
	\end{equation}
where $M_{\rm QAM}$ is the modulation cardinality, $N_{\rm RB} $ is the number of resource blocks, $N_{\rm sc}^{(RB)}$ is the number of subcarriers per resource block, $N_{\rm sc}^{(OFDM)}$ is the number of OFDM symbols per resource block, $\tau_{\rm data}$ is the transmit delay for the data, and $\eta_{\rm CPRI}$ is the efficiency of the Common Public Radio Interface (CPRI)\footnote{Numerical values used for these parameters in the numerical results will be reported later in Section \ref{Performance_Evaluation}.\ref{Simulation_parameters_section}.}.
The data-rate related to the \textit{FH weigths} depends on the number of bits needed to send the beamforming weigths to the $n$-th AP/BS. Note that this value is zero in the case of local processing since in this case the beamforming weights are evaluated locally at the AP/BS and thus they do not need to be transferred over the fronthaul link.
We have:
\begin{equation}
	\text{F}_n^{(W)} = \ds \sum_{k=1}^K \widetilde{a}_{k,n} 2 \frac{N_{\rm RB}}{N_{\rm C,B}^{(RB)}} \frac{N_{\rm T}^{(n)}}{\tau_{\rm weight}} \frac{N_{\rm Q}}{\eta_{\rm CPRI}}, \end{equation}
where the ``2'' takes into account the transmission of the real and imaginary part of the beamforming coefficients, $N_{\rm C,B}^{(RB)}$ is the beamforming granularity (i.e., the number of subcarriers that use the same beamformer), $N_{\rm T}^{(n)}$ is the number of antennas at the $n$-th AP/BS, $\tau_{\rm weight}$ is the transmit delay of the weights and $N_{\rm Q}$ is the bit-width, i.e. the number of quantization bits.
The total FH requirement at the $n$-th AP/BS can be thus written as
\begin{equation}
	\begin{array}{lll}
\text{F}_n=\ds\sum_{k=1}^K  \widetilde{a}_{k,n} \frac{N_{\rm RB}}{\eta_{\rm CPRI}} &\left[ \log_2 (M_{\rm QAM}) \frac{N_{\rm sc}^{(RB)}}{\tau_{\rm data}} N_{\rm sc}^{(OFDM)}\right. \\ & \left. +  2 \frac{N_{\rm Q}}{N_{\rm C,B}^{(RB)}} \ds \frac{N_{\rm T}^{(n)}}{\tau_{\rm weight}} \right] \; .
\end{array}\end{equation}
We now present the proposed heuristic and deterministic algorithm to fulfill the FH requirement in \eqref{eq:FH_requirement}. The procedure is as follows.
\begin{itemize}
	\item[s1)] Initialize the procedure by creating an association of APs and BSs to the users assuming that each UE is served by the best $\widetilde{N}_{\rm AP}$ APs and $\widetilde{N}_{\rm AP}$ BSs. This association is encoded in the coefficients $\widetilde{a}_{k,n}, \; \forall k,n$, with $\widetilde{a}_{k,n}=1$ if the $k$-th user is served by the $n$-th AP/BS and 0 otherwise.	
	\item[s2)] Compute the precoders. Specifically, in the case of local processing discussed in Section \ref{System_design}.\ref{Local_precoding_Section} design the local beamforming at the APs/BSs; in the case of joint processing presented in Section \ref{Joint_PZF_Section}, design the coordinate beamforming at the CPU to be shared over the fronthaul link. This procedure is performed on every $N_{C,B}^{(RB)}$ subcarrier using the averaged channels on consecutive $N_{C,B}^{(RB)}$ subcarriers.
	\item[s3)] Evaluate the FH load $\text{F}_n, \; \forall n$ and define $\mathcal{N}_{\rm FH}=\lbrace n \, : \, \text{F}_n > \bar{F}_n\rbrace$, i.e., select the APs/BSs whose FH requirement exceeds the FH limitation;	
	\item[s4)] For each $\overline{n} \in \mathcal{N}_{\rm FH}$,  define $\mathcal{K}_{\overline{n}}$ as the set of users served by the $\overline{n}$ AP/BS;
     \item[s5)] Compute the following metric for all the users $k \in  \mathcal{K}_{\overline{n}}$ associated to AP/BS $\overline{n}$:
	\begin{equation}
	S_{k,\overline{n}}= \ds \frac{\ds \sum_{\substack{n=1 \\ n \neq \overline{n}}}^{M+L}  \widetilde{a}_{k,n} \delta_{k,n}}{ \ds  \sum_{\substack{j=1 \\ j \neq k}}^K \ds  \sum_{\substack{n=1 \\ n \neq \overline{n}}}^{M+L}  \widetilde{a}_{j,n} \delta_{k,n}+\sigma^2_z}\; ,
	\end{equation}
	where $\delta_{k,n}$ is the large-scale fading coefficient between the $k$-th user and the $n$-th AP/BS. 
	Notice that the metric $S_{k,\overline{n}}$ is a proxy (i.e., an approximation) of the $k$-th UE SINR for the case in which is is no longer served by the $n$-th AP/BS.
	\item[s6)] For each $\overline{n} \in \mathcal{N}_{\rm FH}$ find 
	\begin{equation}
	k^*= \arg \max_{k \in \mathcal{K}_{\overline{n}}}   S_{k,\overline{n}}
	\end{equation}
	and set $\widetilde{a}_{k^*,\overline{n}}=0$; otherwise stated, we are removing the link between AP/BS $\overline{n}$ and the user that after this removal will have the largest value of proxy SINR. This will weaken the FH requirement for AP/BS $\overline{n}$.	
	\item[s7)] Go to s2) and continue until $\mathcal{N}_{FH}=\emptyset$.
\end{itemize}

\begin{figure*}[]
	\centering
	\includegraphics[scale=0.75]{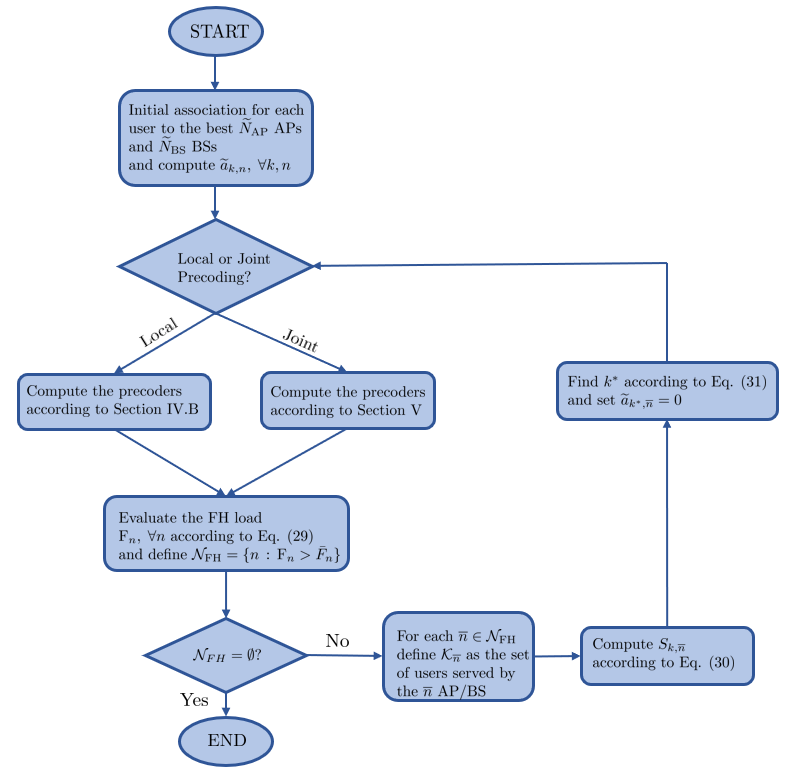}
	\caption{{Flowchart of the precoding design and user association considering limited-capacity fronthaul constraints.}}
	\label{Fig:flowchart}
\end{figure*}

{Figure \ref{Fig:flowchart} presents the detailed flow chart for the procedure mentioned earlier and described in Algorithm \ref{Alg:User_Ass_Fronthaul}.}
 
\begin{algorithm}
	
	\caption{User association with FH requirements}
	
	\begin{algorithmic}[1]
		\label{Alg:User_Ass_Fronthaul}
		\STATE Start with the user association with no constraints on the fronthaul link and define $\widetilde{a}_{k,n} \forall k,n$.
		
		\REPEAT 
		
		\STATE Design the beamforming according to the local or joint processing.
		
		\STATE Evaluate $\text{F}_n, \; \forall n$ and define $\mathcal{N}_{\rm FH}=\lbrace n \, : \, \text{F}_n > \bar{F}_n\rbrace$.
		
		\STATE For each $\overline{n} \in \mathcal{N}_{\rm FH}$ define $\mathcal{K}_{\overline{n}}$ and $S_{k,\overline{n}} \, \forall \, k \in \mathcal{K}_{\overline{n}}$.
		
		\STATE For each $\overline{n} \in \mathcal{N}_{\rm FH}$ find 
		
		\begin{equation*}
		k^*= \arg \max_{k \in \mathcal{K}_{\overline{n}}}   S_{k,\overline{n}}
		\end{equation*}
		
		and set $\widetilde{a}_{k^*,\overline{n}}=0$; 
		
		\UNTIL $\mathcal{N}_{FH}=\emptyset$  
	\end{algorithmic}
	
\end{algorithm}

\section{Numerical Results} \label{Performance_Evaluation}
We now provide simulation results in order to carry out a compared performance analysis of the considered network deployments and proposed algorithms.

\subsection{Simulation setup and parameters} \label{Simulation_parameters_section}
A cluster of three cells surrounded by nine cells is considered to take into account the effect of intercell interference. 
Each cell is actually made of three sectors with $2\pi/2$ coverage; each of these sectors is treated as an independent base station and is equipped with a ULA with half-wavelength spacing. The positions of the users are uniformly generated in each cell in the annulus with minimal radius (that is, distance from a BS) 15~m and maximal radius set at $0.97  \text{ISD}/2$, where $\text{ISD}$ is the ``inter-site distance'', i.e. the distance between BSs. We consider two distributions for the APs: uniformly distributed and cell-edge placement. In the former case, the APs are randomly and uniformly distributed similarly to the users, while in the latter scenario the APs are placed on a circle with radius $0.8 \, \text{ISD}/2$. In Fig. \ref{Fig:scenario} we report a snapshot of the 2D view of the scenario considered with $\text{ISD}=500$~m.

With regard to the model for the large-scale fading coefficients, we use the model in reference \cite{3GPP_38901_model}.
{We start by describing the channel between the BSs and the users, which are modeled according to the urban macrocell (UMa) model, using the parameters reported in \cite[Table 7.4.1-1]{3GPP_38901_model}. In order to discriminate between LOS and non-line-of-sight (NLOS) model, we firstly evaluate the LOS probability according to the parameters reported in in \cite[Table 7.4.2-1]{3GPP_38901_model}. For the UMa scenario, the LOS probability depends on the 2D distance between the $k$-th user and the $\ell$-th BS, $d_{k,\ell}^{(b)}$, according to the expression in Eq. \eqref{P_LOS_BS} at the top of next page, where $h_{k}$ is the antenna height at the $k$-th user, and
\begin{equation}
    C\left(h_{k}\right) = \left \lbrace 
    \begin{array}{cc}
     0 & \quad \text{if} \; h_k \leq 13 \, \text{m} \\
     \left(\ds \frac{h_k-13}{10}\right)^{1.5}  & \quad \text{if} \; 13 < h_k \leq 23 \, \text{m} 
    \end{array}\right. .
    \label{C_h_k}
\end{equation}}

\begin{figure*}
{
\begin{equation}
    p_{\rm LOS}^{(b)}\left(d_{k,\ell}^{(b)}\right)=\left \lbrace \begin{array}{cc}
           1 & \quad \text{if} \; d_{k,\ell}^{(b)} \leq 18 \, \text{m} \\
          \left[ \ds \frac{18}{d_{k,\ell}^{(b)}} + \left( 1- \ds \frac{18}{d_{k,\ell}^{(b)}} \right) e^{-\frac{d_{k,\ell}^{(b)}}{63}}\right] \left( 1 + \ds \frac{5}{4}C\left(h_{k}\right) \left( \frac{d_{k,\ell}^{(b)}}{100}\right)^3 e^{-\frac{d_{k,\ell}^{(b)}}{150}} \right)  & \quad \text{otherwise}
    \end{array}\right.  
    \label{P_LOS_BS}
\end{equation}}
\end{figure*}

\begin{figure}[H]
	\centering
	\includegraphics[scale=0.45]{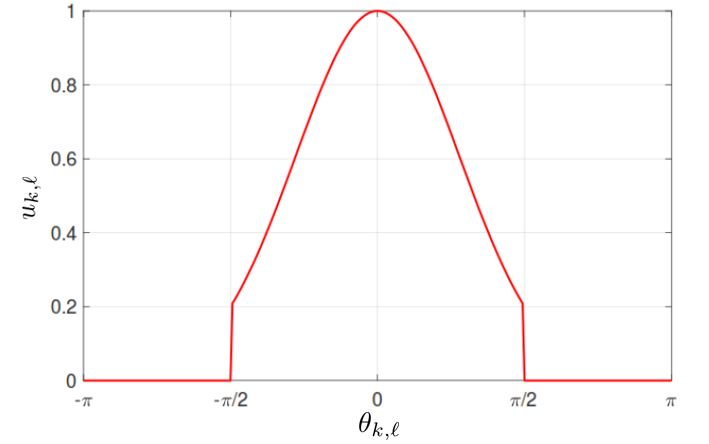}
	\caption{Array gain versus $\theta_{k,\ell}$, the angle between the $k$-th user and the $\ell$-th BS.}
	\label{Fig:array_gain}
\end{figure}

Then, recalling Eq. \eqref{channel_model_b}, the large-scale fading coefficient between the $k$-th user and the $\ell$-th BS in is $\rho_{k,\ell}=u_{k,\ell}\widetilde{\rho}_{k,\ell}$. Here, $u_{k,\ell}$ takes into account the  gain of the ULA array in the BS. We assume that the array gain is modeled with a Gaussian shaping radiating onto a $[ -\pi/2, \pi/2]$ angular region as reported in Fig. \ref{Fig:array_gain}.
The large-scale fading coefficient $\widetilde{\rho}_{k,\ell}$ in logarithmic unit is evaluated as follows
\begin{equation}
	\widetilde{\rho}_{k,\ell} [\text{dB}]=-\text{PL}_{k,\ell}^{(b)}+z_{k,\ell}^{(b)},
\end{equation} 
where $\text{PL}_{k,\ell}^{(b)}$ represents the path loss and $z_{k,\ell}^{(b)}$ represents the shadow fading. 
Details on the path loss modeling are reported in the Appendix.
The shadow fading coefficients from a BS to different users are correlated and follow \cite[Eq. (7.4-5)]{3GPP_38901_model} 
\begin{equation}
	\mathbb{E}[z_{k,\ell}^{(b)}z_{j,t}^{(b)}]= \left\lbrace
	\begin{array}{llll}
		\sigma_{(b),{\rm SF}}^2 e^{-{\bar{d}_{k,j}^{(b)}}/{d_{\rm corr}^{(b)}}} \;, & \ell=t, \\
		0 \; ,& \ell \neq t \, ,
	\end{array} \right.
	\label{shadowing_correlation_b}
\end{equation}
where $\bar{d}_{k,j}^{(b)}$ is the distance between the $k$-th and the $j$-th outdoor users and $d_{\rm corr}^{(b)}=50$m is the correlation distance and follows \cite[Table 7.5-6 Part-1]{3GPP_38901_model} and $\sigma_{(b),{\rm SF}}=6$ \cite[Table 7.4.1-1]{3GPP_38901_model}.

For the channels between the APs and the users, we use the urban microcell (UMi) model. Again,in order to discriminate between LOS and NLOS model, we first evaluate the probability of LOS according to the parameters reported in \cite[Table 7.4.2-1]{3GPP_38901_model}. For the UMi scenario, the LOS probability depends on the 2D distance between the $k$-th user and the $m$-th AP, $d_{k,m}^{(a)}$, according to the expression in Eq. \eqref{P_LOS_AP} at the top of next page.

\begin{figure*}
{
\begin{equation}
    p_{\rm LOS}^{(a)}\left(d_{k,m}^{(a)}\right)=\left \lbrace \begin{array}{cc}
           1 & \quad \text{if} \; d_{k,m}^{(a)} \leq 18 \, \text{m} \\
           \ds \frac{18}{d_{k,m}^{(a)}} + \left( 1- \ds \frac{18}{d_{k,m}^{(a)}} \right) e^{-\frac{d_{k,m}^{(a)}}{36}}  & \quad \text{otherwise}
    \end{array}\right.  
    \label{P_LOS_AP}
\end{equation}}
\hrulefill
\end{figure*}

Then, the path-loss between the $k$-th user and the $m$-th AP is modeled according to \cite[Table 7.4.1-1]{3GPP_38901_model}.
The large-scale fading coefficient $\beta_{k,m}$ in Eq. \eqref{channel_model_a} is evaluated in logarithmic unit as follows
\begin{equation}
	\beta_{k,m} [\text{dB}]=-\text{PL}_{k,m}^{(a)}+z_{k,m}^{(a)} \, ,
\end{equation} 
where $\text{PL}_{k,m}^{(a)}$ is the path loss (described in the Appendix) and $z_{k,m}^{(a)}$ represents the shadow fading. The shadow fading coefficients from an AP to different users are correlated with an expression similar to \eqref{shadowing_correlation_b}, where the values $\bar{d}_{k,j}^{(b)}$ now equal 13m, and $\sigma_{(b), {\rm SF}}=7.82$.

The chosen carrier frequency is $f_c=3.5$ GHz, the system bandwidth is $B=20$ MHz, the power spectral density of the noise is  $N_0=-174$ dBm/Hz and the noise figure at the receiver is $F=9$ dB. We consider in the whole simulation area $Q=12$ macro BSs with three sectors, i.e., the total number of BSs is $L=36$, the number of APs is $M=108$, 9 in each cell. We equip the BSs ULA with 32 antennas and the APs with 8 antennas arrays; the height of the BSs is $h_{\rm BS}=25$~m and the height of the APs is $h_{\rm AP}=10$~m. We assume $\tau_c =640$ time/frequency samples, corresponding to a coherence bandwidth of $640$ kHz and a coherence time of $1$~ms. Equal uplink/downlink split of the available time/frequency resources after training is assumed, i.e., $\tau_{\rm d}=\frac{\tau_c-\tau_p}{2}$. We take a set $\mathcal{P}_{\tau_p}$ of orthogonal pilots with length $\tau_p=32$ and the pilot sequences in $\mathcal{P}_{\tau_p}$ are assigned to the users according to the $k$-means Algorithm discussed in Section \ref{System_model}.\ref{Pilot_Assignment_Algorithm}, i.e., the results account for the impact of \textit{pilot contamination}. The uplink transmit power during training is $\eta_k=\tau_p \overline{\eta}_k$, with $\overline{\eta}_k=300$~mW $ \forall k=1,\ldots,K$. We present results for the case of LMMSE channel estimation discussed in Section \ref{System_model}.\ref{LMMSE_channel_estimation}. All the system parameters are summarized in Table \ref{table:parameters}. 
\begin{table}[]
	\centering
	\caption{System parameters}
	\label{table:parameters}
	\def\arraystretch{1.2}
	\begin{tabulary}{\columnwidth}{ |p{1.1cm}|p{4.5cm}|p{1.8cm}| }
		\hline
		\textbf{Name} 				& \textbf{Meaning} & \textbf{Value}\\ \hline
		$M$ 				& number of total APs in the simulation area & 108\\ \hline
		$L$ 				& number of total BSs (sectors) in the simulation area & 36\\ \hline
		$N_{\rm BS}$ 				& number of antennas at the BSs& 32\\ \hline
		$N_{\rm AP}$ 				& number of antennas at the APs& 8\\ \hline
		AP dist.				&  Distribution of APs in the system & Uniform or Cell-Edge placement\\ \hline
		MS dist.				&  Distribution of MSs in the system & Uniform placement\\ \hline
		$P_{\ell,{\rm max}}^{(b)}$	&   maximum power available at the $\ell$-th BS & 46 dBm\\ \hline
		$P_{m,{\rm max}}^{(a)}$	&   maximum power available at the $m$-th AP & 39 dBm\\ \hline
        $\widetilde{N}_{\rm BS}$ & number of BSs serving each user in the FULL-Coop scenario  &  3\\ \hline 
        $\widetilde{N}_{\rm AP}$ & number of APs serving each user in the FULL-Coop scenario & 6\\ \hline
        $N_{\rm PZF}^{\rm BS}$ & number of users to be protected by the interference generated by each BS & $N_{\rm BS}/2=16$\\ \hline
        $N_{\rm PZF}^{\rm AP}$ & number of users to be protected by the interference generated by each AP &$N_{\rm AP}/2=4$ \\ \hline
        $R_{\rm JPZF}$ & number of users to be protected from the interference generated by the downlink transmission to the generic user in the centralized precoding scheme  & $\begin{array}{lll} \!\!\!\!\!\!\left(N_{\rm AP} \widetilde N_{\rm AP} + \right.\\  \!\!\!\!\!\!\!\!\!\left. N_{\rm BS} \!\widetilde N_{\rm BS}\right)\!\!/2\!=\!72\end{array}$ \\ \hline
		$f_c$	&  Carrier frequency & 3.5 GHz\\ \hline
		$B$	&  Bandwidth & 20 MHz\\ \hline
		$N_0$			& power spectral density of the thermal noise & -174 dBm/Hz\\ \hline
		$F$ 			& Noise figure at the receiver & 9 dB\\ \hline	
		 $M_{\rm QAM}$ 			& modulation cardinality & 256\\ \hline	
		 $N_{\rm RB}$ 			& number of resource blocks & 55\\ \hline	
		 $N_{\rm sc}$ 			& number of subcarriers & 1024\\ \hline	
		  $N_{\rm sc}^{(RB)}$ 			& number of subcarriers per resource blocks & 19\\ \hline	
		  $N_{\rm sc}^{(OFDM)}$ 			& number of OFDM symbols per subcarrier & 14\\ \hline	
		   $\tau_{\rm data}$ 			& transmit delay for the data & 0.5 ms \\ \hline
		   $\tau_{\rm weight}$ 			& transmit delay for the weights & 0.2 ms \\ \hline
		     $N_{\rm Q}$ 			&  bit-width &  8 \\ \hline
		     $N_{\rm C,B}^{(RB)}$ &  beamforming granularity &  64 \\ \hline
		   $\eta_{\rm CPRI}$ 			& efficiency of the Common Public Radio Interface (CPRI) & 0.85 \\ \hline
		   $\bar{F}_n \; \forall n$ & fronthaul load limit for all the APs and BSs & 5 Gbps \\ \hline
	\end{tabulary}
\end{table}

The system performance is represented in terms of the cumulative distribution function (CDF) of the downlink rate, evaluated using the upper bound expression of the spectral efficiency reported in Eq. \eqref{SE_DL_ICSI_co-existence_UpperBound} for cell-inside and cell-edge users. 

\begin{figure*}[t]
	\centering
	\includegraphics[scale=0.45]{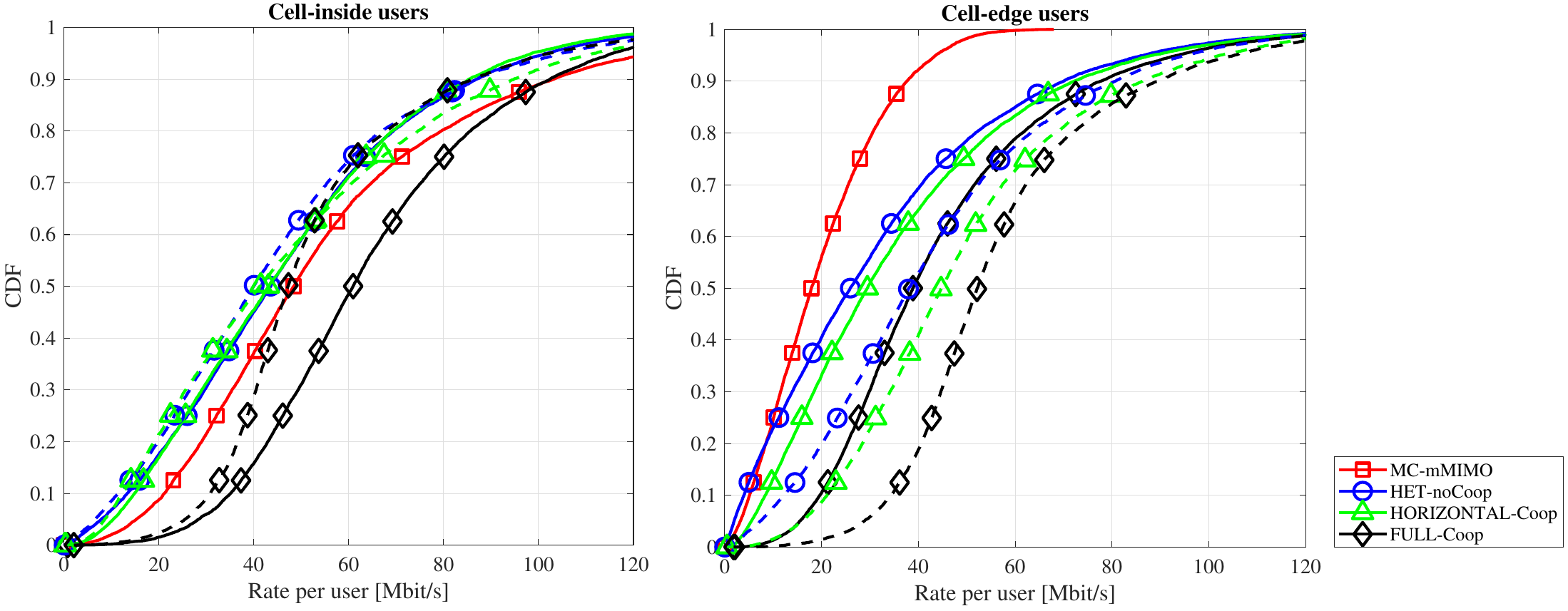}
	\caption{CDF of the rate per user in the four considered scenarios with FPA with $\alpha=-0.5$, MMSE beamfoming at the BSs and APs and 5 users per sector, i.e., $K=180$. Continuous lines represent uniform APs placement in the cells and dashed lines represent cell-edge APs placement.}
	\label{Fig:comparison4scenarios}
\end{figure*}

\subsection{Comparison of the four considered scenarios}

{Given that the majority of the existing literature on CF-mMIMO systems predominantly addresses the Rayleigh model for small-scale fading, we begin by concentrating on this model, specifically assuming $K_{k,m}^{(a)}=K_{k,\ell}^{(b)}=0$, and defer the discussion of Rician fading to Section \ref{Performance_Evaluation}.\ref{Num_results_Rician}.}
In Figure \ref{Fig:comparison4scenarios}, we present the CDF illustrating the downlink rate per user across four examined scenarios utilizing FPA with a parameter of $\alpha=-0.5$, alongside MMSE beamforming at both BSs and APs, accommodating five users per sector (i.e., $K=180$). The plot on the left delineates the performance of users within the cell, while the plot on the right depicts the performance of cell-edge users. Solid lines denote uniform AP placement, while dashed lines denote cell-edge AP placement.
Upon examination of the figure, several observations emerge. Firstly, regarding the deployment of MC-mMIMO, the plots corroborate the significant discrepancy in user performance between those situated within the cell and those positioned at the cell-edge. Although MC-mMIMO configurations excel in providing remarkable performance to nearby users, they encounter interference limitations when serving users located at the cell-edge. Additionally, as anticipated, the FULL-Coop scenario yields the most favorable overall performance. However, it should be noted that there exists an exception in which MC-mMIMO outperforms FULL-Coop for cell-interior users in the upper right segment of the CDF. This can be attributed to the detrimental impact of cooperation for users in close proximity to a massive MIMO array, as the additional diversity links are inferior to the direct link with the nearby massive MIMO antenna array.
Furthermore, among the schemes exhibiting intermediate performance, HET-noCoop and HORIZONTAL-Coop stand out, with the latter marginally surpassing the former. Regarding AP placement strategies, the results indicate that placing APs at the cell-edge proves advantageous for cell-edge users but disadvantageous for cell interior users. This outcome aligns with expectations, as cell-edge AP placement confers benefits to users on the cell-edge while somewhat impeding those within the cell. Nevertheless, the degradation experienced by cell-interior users pales in comparison to the gains accrued by cell-edge users, reaffirming the intuition that placing APs at the cell-edge is preferable.
It is also worth noting that, although not explored in this study, the use of a more sophisticated association rule between UEs and APs/BSs may mitigate the performance degradation for cell-interior users. This aspect is left for further investigation in future research endeavors.

\subsection{Impact of the beamformers}
\begin{figure*}[]
	\centering
	\includegraphics[scale=0.4]{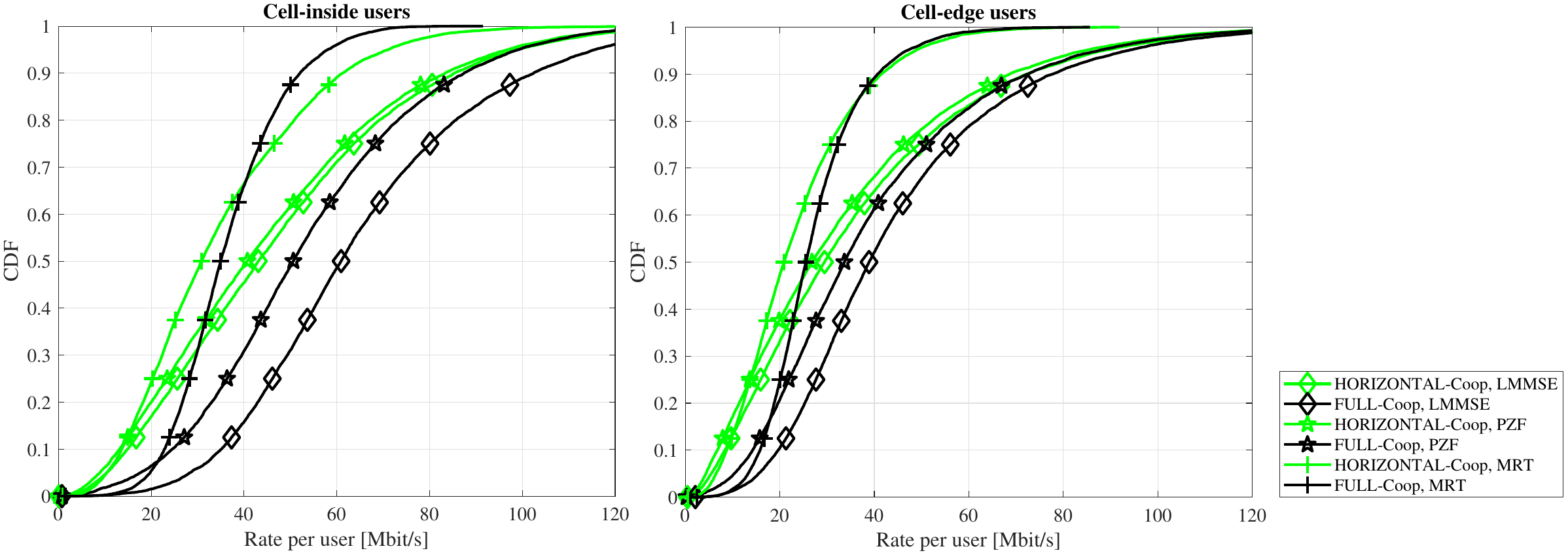}
	\caption{CDF of the rate per user in the HORIZONTAL-Coop and FULL-Coop scenarios with FPA with $\alpha=-0.5$, 5 users per sector, i.e., $K=180$, and uniform APs distribution. Comparison between the beamforming techniques at the APs and BSs.}
	\label{Fig:comparison_beamforming}
\end{figure*}

In Figure \ref{Fig:comparison_beamforming}, our focus is on the two most promising deployments, namely the HORIZONTAL-Coop and FULL-Coop scenarios, where we present the CDF delineating the downlink rate per user employing FPA with $\alpha=-0.5$. In this analysis, again we assume five users per sector (i.e., $K=180$), and a uniform distribution of APs.
The results affirm the well-established hierarchy among the considered beamformers, showcasing that MMSE achieves superior performance, followed by PZF, and, lastly, MRT combining. Notably, the FULL-Coop deployment surpasses the HORIZONTAL-Coop, with one notable exception observed in the upper portion of the CDF for cell-interior users when MRT is employed.
This anomaly can be explained by the limited capacity of MRT beamforming to mitigate interference. In scenarios involving cell-interior users, coercing cooperation with APs and other BSs, as observed in the FULL-Coop scenario, leads to an overall increase in interference. While MMSE and PZF beamformers manage this escalation well, MRT does not, thereby rendering connecting to other APs/BSs disadvantageous for users benefiting from optimal channel conditions from nearby arrays. This effect could potentially be alleviated by devising more intelligent association rules between UEs and APs/BSs.
Moreover, in the lower segment of the CDF, it is discernible that MRT marginally outperforms PZF beamforming, perhaps attributable to the noise amplification effect resulting from projections onto reduced-dimensionality subspaces.
However, in general, it is evident that the FULL-Coop deployment, employing either MMSE or PZF beamforming, achieves the most commendable performance across all examined scenarios.

\subsection{Impact of the power allocation strategies}

\begin{figure*}[]
	\centering
	\includegraphics[scale=0.4]{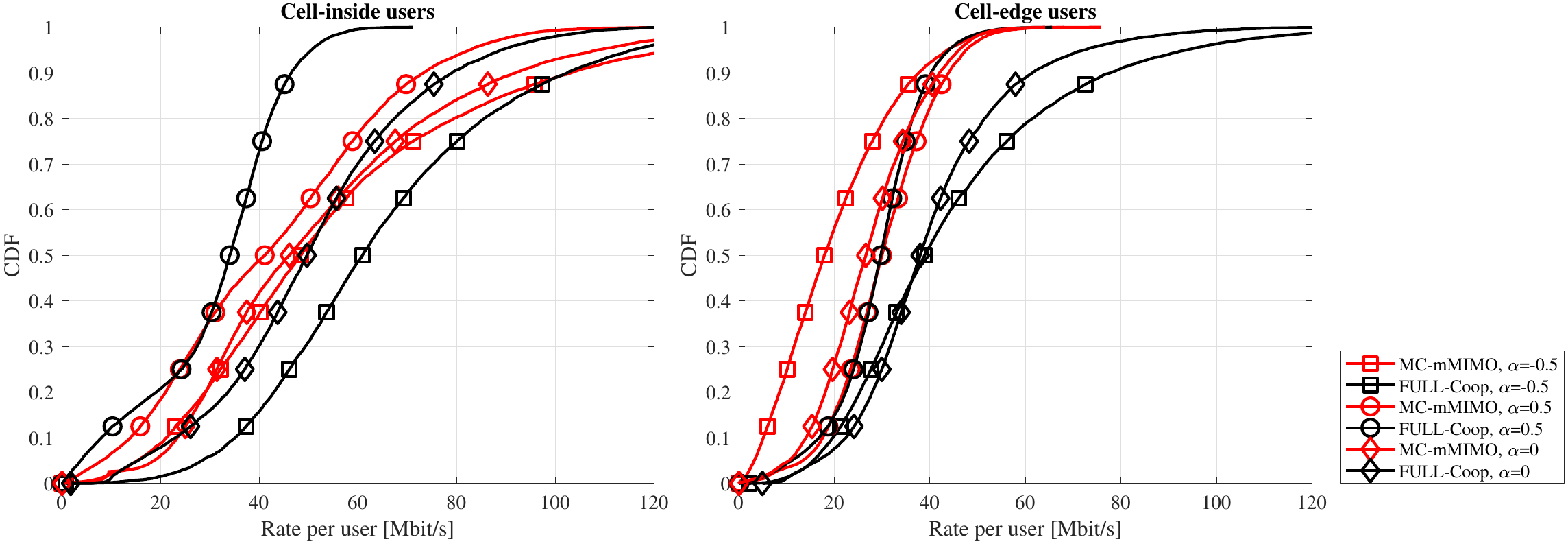}
	\caption{CDF of the rate per user in the MC-mMIMO and FULL-Coop scenarios with three values of $\alpha$ in the FPA strategy, MMSE beamforming, uniform APs distribution and 5 users per sector, i.e., $K=180$.}
	\label{Fig:comparison_power_allocation}
\end{figure*}

In Figure \ref{Fig:comparison_power_allocation}, we examine the influence of the power allocation strategy on the performance of the system. Our analysis is centered on the MC-mMIMO and FULL-Coop scenarios, assuming MMSE beamforming, a uniform distribution of APs, and 5 users per sector, i.e. we have $K=180$. The figure illustrates the cumulative distribution function (CDF) of the rate per user under the Fixed Power Allocation (FPA) scheme specified in \eqref{FPA_powers}, with three different values of $\alpha$: $\alpha= \pm 0.5$ and $\alpha=0$. Notably, $\alpha=0$ corresponds to uniform power allocation, $\alpha=0.5$ represents partial channel inversion aimed at reducing performance discrepancies among users, while $\alpha=-0.5$ favors network throughput by allocating more power to users with superior channel conditions. Similarly to previous figures, the performance of users within the cell and those at the cell periphery is illustrated separately in two distinct plots. Insightful observations can be made by examining these plots. For the MC-mMIMO scenario, the choice $\alpha=-0.5$ provides the best performance for the cell-inside users (those getting the best channel conditions) and the worst performance for the users at the cell-edge. This experimental result confirms that $\alpha=-0.5$ aims to maximize network throughput, neglecting weak users at the expense of network fairness. For the FULL-Coop scenario, instead, a different behavior is obeserved; precisely, the choice $\alpha=-0.5$ provides by far the best performance for cell-inside users, but it also behaves well for the users at the cell-edge. This behavior can be explained by noticing that, thanks to full cooperation, macro-diversity gains provide stable and reasonably good channel conditions to all the users in the system, and, thus, there is no need to make conservative choices (i.e., taking $\alpha=0.5$) to promote network fairness.

\begin{figure*}[t]
	\centering
	\includegraphics[scale=0.42]{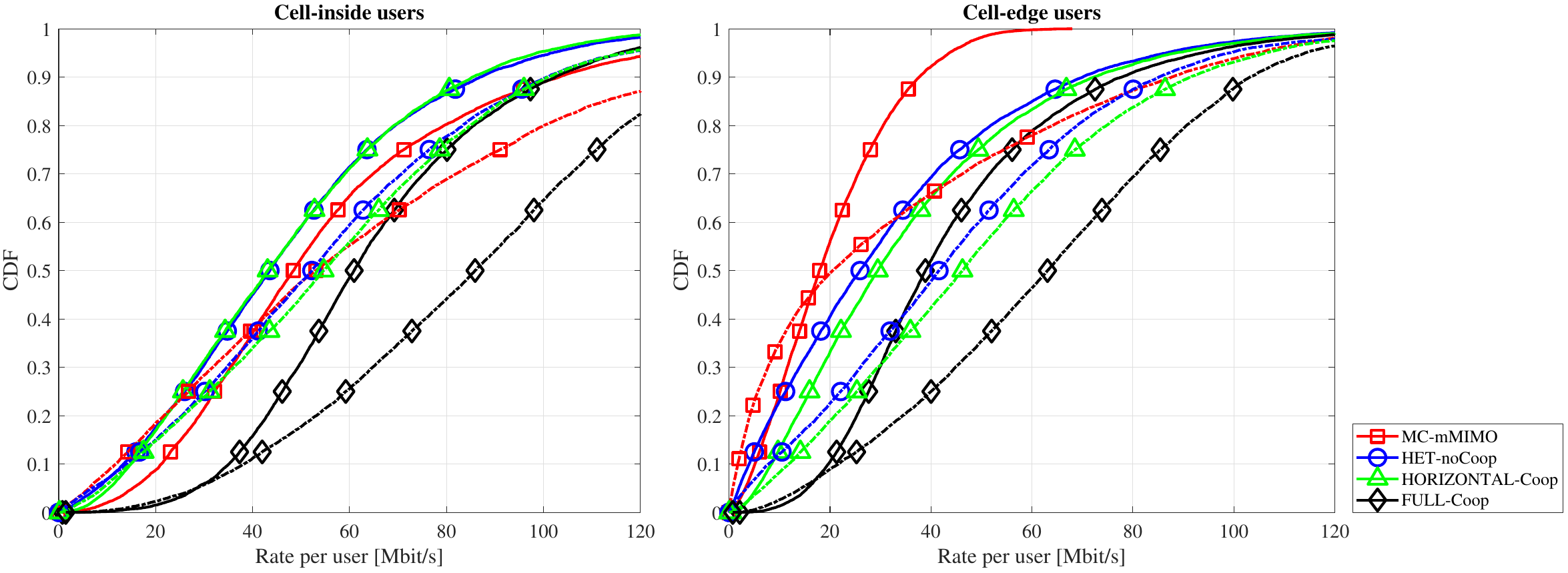}
	\caption{ CDF of the rate per user in the four considered scenarios with FPA with $\alpha=-0.5$, MMSE beamfoming at the BSs and APs, 5 users per sector, i.e., $K=180$, and uniform APs placement. Continuous lines represent Rayleigh fading (i.e., $K_{k,m}^{(a)}=K_{k,\ell}^{(b)}=0$) and dash-dotted lines represent Rician model detailed in the paper, following the 3GPP 38.901 model.}
	\label{Fig:comparisonRayleigh_Rice}
\end{figure*}

{
\subsection{Impact of Rician-distributed small-scale fading} \label{Num_results_Rician}}
{In Figure \ref{Fig:comparisonRayleigh_Rice}, we analyze the performance of the four considered cooperation/co-existence scenarios considering the Rician and Rayleigh channel models described in Section \ref{System_model}.\ref{Channel_model}.  We assume FPA with a parameter of $\alpha=-0.5$, alongside MMSE beamforming at both BSs and APs, accommodating five users per sector (i.e., $K=180$) and uniform placement of APs. Solid lines denote the performance with the Rayleigh channel model, i.e., $K_{k,m}^{(a)}=K_{k,\ell}^{(b)}=0$, while dash-dotted lines denote the performance with Rician fading, where $K_{k,m}^{(a)}$ and $K_{k,\ell}^{(b)}$ depend on the LOS probability following the model reported in \cite{3GPP_38901_model}.  
Examining the figures, it is seen that the relative ranking of the four considered cooperation scenarios does not significantly change; however, the values of the downlink rate when a LOS path is present are larger than those obtained when the LOS path is obstructed. By observing the performance of the MC-mMIMO, we can see that in the lower-left part of the CDF, i.e., the one referring to the worst users, the Rician fading case is outperformed by the Rayleigh one. This behavior can be justified by the higher interference that the worst users experience from the nearby BSs that do not cooperate with the serving one to provide better coverage. In the FULL-Coop scenario, conversely, the Rician fading is always beneficial from the user perspective, but if we focus on the 95\%-likely user performance, the two channel models offer very similar performance. 
Overall, it can be concluded that Rician-distributed channels achieve in most cases better performance than Rayleigh-distributed channels, but the difference is not relevant when focusing on the 95\%-likely user performance. Moreover, the relative ranking of the cooperation scenarios considered is mainly preserved. 
 For these reasons, and particularly because rich scattering environments and Rayleigh fading channels serve as the standard benchmark for most simulations in this area of research, the remaining performance results in this study consider the case $K_{k,m}^{(a)}=K_{k,\ell}^{(b)}=0$.}

\subsection{Impact of the centralized precoding schemes and of the limitation of the fronthaul load}
\begin{figure*}[]
	\centering
	\includegraphics[scale=0.45]{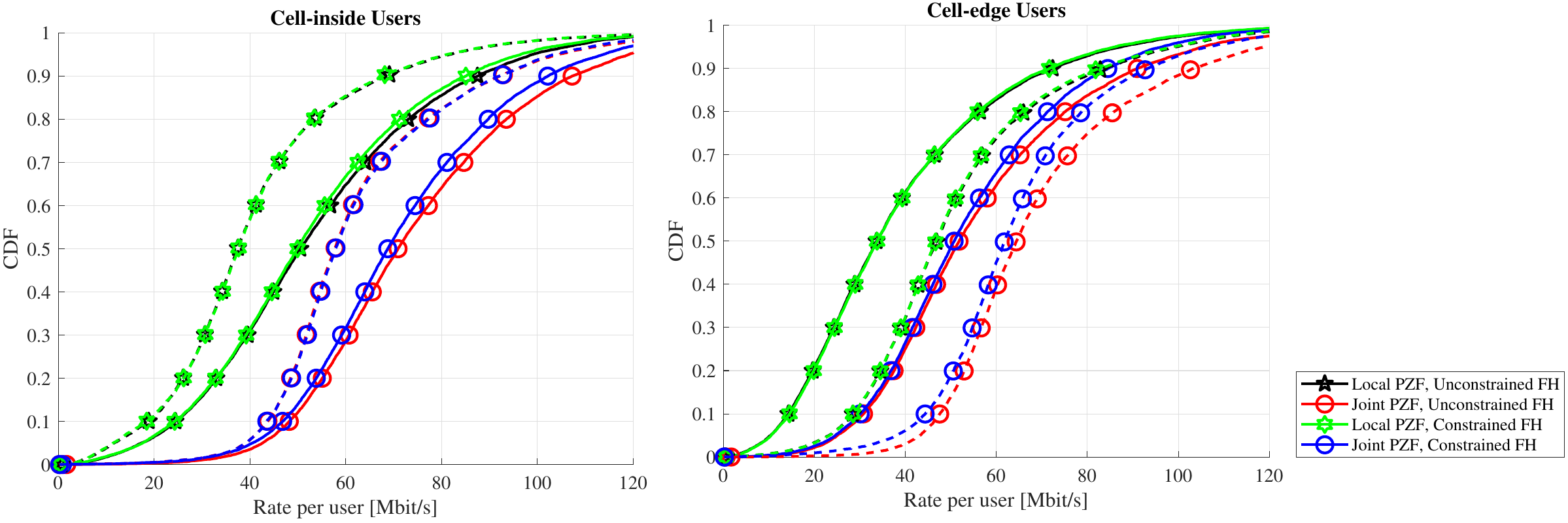}
	\caption{CDF of the rate per user in the FULL-Coop scenario with 5 users per sector, i.e., $K=180$, FPA with $\alpha=-0.5$. Impact of the joint processing and of the limitation of the fronthaul load. Continuous lines represent uniform APs placement in the cells and dashed lines represent cell-edge APs placement.}
	\label{Fig:comparison_FH_constraints_Ks_5}
\end{figure*}

\begin{figure*}[]
	\centering
	\includegraphics[scale=0.45]{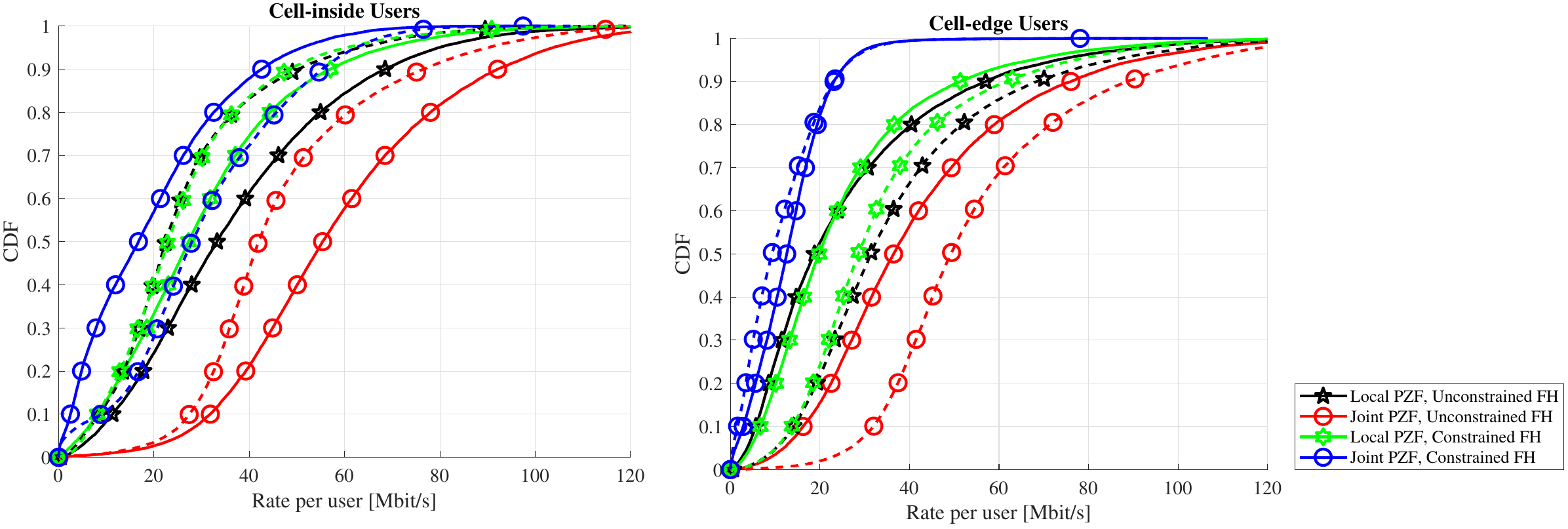}
	\caption{CDF of the rate per user in the FULL-Coop scenario with 9 users per sector, i.e., $K=324$, FPA with $\alpha=-0.5$. Impact of the joint processing and of the limitation of the fronthaul load. Continuous lines represent uniform APs placement in the cells and dashed lines represent cell-edge APs placement.}
	\label{Fig:comparison_FH_constraints_Ks_9}
\end{figure*}

We proceed to examine the influence of the centralized precoding scheme compared to the local precoding scheme, simultaneously evaluating the impact of enforcing compliance with FH requirements on system performance.
In Figures \ref{Fig:comparison_FH_constraints_Ks_5} and \ref{Fig:comparison_FH_constraints_Ks_9}, we present the CDF of the rate per user within the FULL-Coop scenario using FPA with $\alpha=-0.5$, considering cases with 5 and 9 users per sector, respectively. Thus, the total number of users in the system amounts to $K=180$ for Figure \ref{Fig:comparison_FH_constraints_Ks_5} and $K=324$ for Figure \ref{Fig:comparison_FH_constraints_Ks_9}. Our analysis distinguishes between the performance of users located within the cell and those at the cell-edge. We employ the PZF beamformer and consider both uniform and cell-edge AP placements.
Upon reviewing Figure \ref{Fig:comparison_FH_constraints_Ks_5}, it becomes evident that the joint PZF beamformer notably outperforms the local PZF beamformer, aligning with our anticipated expectations. Furthermore, the adherence to the FH requirements introduces a marginal performance degradation in the case of joint PZF beamforming, while the impact on performance in the case of local PZF beamforming is negligible and not discernible in the plots.

Contrasting considerations emerge from Figure \ref{Fig:comparison_FH_constraints_Ks_9}, which represents the performance of a highly loaded system. Here, it is observed that meeting the FH constraint results in a noticeable performance decline. However, this degradation is more pronounced in scenarios employing joint centralized beamforming computations compared to those employing local beamforming computations. This discrepancy is rationalized by the fact that beamformers computed at the Central Processing Unit (CPU) necessitate transfer over the FH to the APs/BSs, thereby imposing larger requirements on the FH. Consequently, a greater number of UEs must be disconnected from the APs/BSs to comply with these requirements, resulting in the observed performance differences. Consequently, the joint PZF beamformer exhibits superior performance over the local PZF beamformer only in unconstrained FH scenarios. Conversely, when considering the limited data rate available on the FH links, it emerges as the least favorable solution. This is a phenomenon that, although counterintuitive, finds justification in the provided explanation. These findings suggest the potential for further research and improvements in the field of fronthaul data quantization and the development of effective association rules between UEs and APs/BSs.

\begin{figure*}[]
	\centering
	\includegraphics[scale=0.45]{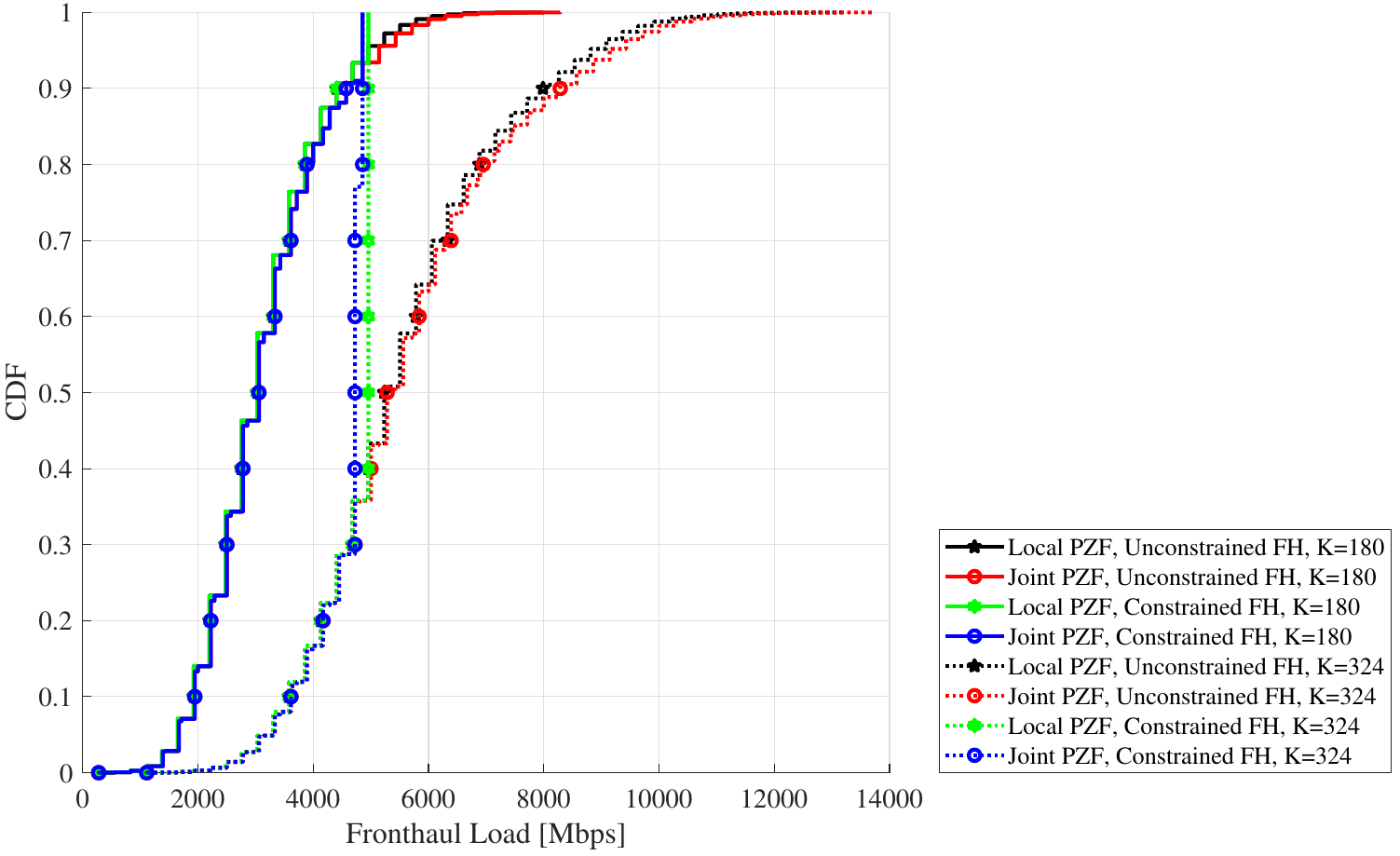}
	\caption{CDF of the fronthaul load in the FULL-Coop scenario with 5 and 9 users per sector, i.e., $K=180$ and $K=324$, FPA with $\alpha=-0.5$. Impact of the joint processing and of the limitation of the fronthaul load.}
	\label{Fig:comparison_FH_loads}
\end{figure*}

Lastly, Figure \ref{Fig:comparison_FH_loads} presents the CDF of data rates on the FH links for both cases: $K=5$ users per sector and $K=9$ users per sector. This figure furnishes valuable insights, illustrating the data rates required on FH links for a CF-mMIMO system with no FH constraints. In particular, FH data rates escalate with the number of active users, with the median data rate on FH links surpassing 5 Gbps for the scenario featuring 9 users per sector. Furthermore, the figure demonstrates that, in constrained FH scenarios, the CDFs exhibit a vertical slope beyond a certain threshold, corresponding to the FH limit. This phenomenon is expected, as the proposed association algorithm terminates some connections between users and APs/BSs until FH limits are satisfied. For the scenario with 5 users per sector, less than 10$\%$ of links exceed FH constraints. In contrast, for the scenario with 9 users per sector, approximately 70$\%$ and 60$\%$ of links violate the FH constraints for joint and local beamforming, respectively. These fairly large values explain the significant performance gap observed between unconstrained and constrained FH solutions for the scenario that features 9 users per sector.

\begin{figure*}
\begin{equation}
		\text{PL}_{k,\ell}^{(b), {\rm LOS}} [\text{dB}]= \left\lbrace
		\begin{array}{llll}
			28.0 + 22\log_{10} \left( d_{k,\ell, {\rm 3D}}^{(b)} \right) + 20\log_{10} (f_{c, {\rm GHz}}) \quad \quad \text{if} \; 10\;\text{m} \leq   d_{k,\ell}^{(b)} \leq  \widetilde{d}_{k,\ell, {\rm BP}}^{(b)} \\
			28.0 + 40\log_{10} \left( d_{k,\ell, {\rm 3D}}^{(b)} \right) + 20\log_{10} (f_{c, {\rm GHz}})  - 9 \log_{10} \left[ \left( \widetilde{d}_{k,\ell, {\rm BP}}^{(b)} \right)^2 + \left( h_{\rm BS} - h_{k}\right)^2 \right] \; \text{if} \;  d_{k,\ell}^{(b)} \geq  5\; \text{km} \\
		\end{array} \right.
	\label{PL_dB_BS}
\end{equation}
\hrulefill
\end{figure*}

\section{Conclusions}\label{Conslucions}
This study has addressed the issue of simultaneous operation and collaboration between traditional MC-mMIMO and the novel CF-mMIMO network deployments. Given the expected gradual deployment of CF-mMIMO architectures in regions where MC-mMIMO network infrastructures are already in place, the study has examined three potential integration scenarios and compared them with a baseline MC-mMIMO system. The study has presented algorithms for downlink beamforming, AP-UE association, power control, and meeting the FH requirement, conducting a comprehensive performance evaluation through extensive numerical simulations. Various AP deployment strategies have been explored, and performance outcomes for UEs located within each cell's core and at the cell periphery have been reported. In general, the findings indicate that increased levels of integration between APs and BSs are generally advantageous, highlighting significant performance discrepancies between users in the cell periphery and those in close proximity to BSs. For the former group, collaboration yields substantially higher benefits compared to the latter. The study has also noted that in certain scenarios, the limitation on FH data rates could offset the advantages derived from centralized beamforming, and that in a fully cooperative network environment, employing power control strategies to enhance fairness among users may not always be optimal. In conclusion, the issue of collaboration and co-existence between BSs and APs is believed to be of significant importance and is likely to attract further research attention in the coming years.

\section*{Appendix: Path loss models}
We report here details on the models used for the UE-BS path loss $\text{PL}_{k,\ell}^{(b)}$ and for the UE-AP path loss $\text{PL}_{k,m}^{(a)}$ according to the model in \cite[Table 7.4.1-1]{3GPP_38901_model}. {For the two cases, we first evaluate the LOS probability according to Eqs. \eqref{P_LOS_BS} and \eqref{P_LOS_AP} and then we select the LOS or NLOS path loss model.}

With regard to the term $\text{PL}_{k,\ell}^{(b)}$, in the case of LOS model the path-loss between the $k$-th user and the $\ell$-th BS is expressed in Eq. \eqref{PL_dB_BS} where $d_{k,\ell, {\rm 3D}}^{(b)}$ and $d_{k,\ell, 
 }^{(b)}$ are the 3D and 2D distances between the $k$-th user and the $\ell$-th BS measured in meters, respectively, $f_{c, {\rm GHz}}$ is the carrier frequency measured in GHz and $ h_{\rm BS} $ and $h_{k}$ are the heights of the BSs and user antennas in meters, respectively.
The breakpoint distance $\widetilde{d}_{k,\ell, {\rm BP}}^{(b)}$ is obtained as
\begin{equation}
	\widetilde{d}_{k,\ell, {\rm BP}}^{(b)}= 4 \ds \frac{h_{\rm BS}^{'}}{h_{{\rm MS},k}^{'}} \frac{f_c}{c},
\end{equation}
with $f_c$ is the carrier frequency in Hz and $c=3 \, 10^{8}$m/s is the light speed in free space, and $h_{\rm BS}^{'}$ and $h_{k}^{'}$ are the effective antenna heights at the BS and the $k$-th user, respectively. The effective antenna heights are calculated as $h_{\rm BS}^{'} = h_{\rm BS}-h_{k,\ell,{\rm E}}^{(b)}$, $h_{k}^{'} = h_{k}-h_{k,\ell,{\rm E}}^{(b)}$,
where $h_{k,\ell,{\rm E}}^{(b)}$ is the effective environment height. The effective environment height is computed as
\begin{equation}
	h_{k,\ell,{\rm E}}^{(b)}=\left \lbrace
	\begin{array}{lll}
		1\;\text{m} & \text{with probability} \; \; p_{k,\ell}^{(b)} \\
		12\;\text{m} & \text{with probability} \;  \; \ds \frac{1-p_{k,\ell}^{(b)}}{3} \\
		 15\;\text{m} & \text{with probability} \; \; \ds \frac{1-p_{k,\ell}^{(b)}}{3} \\
		 h_{k}-1.5\;\text{m} & \text{with probability} \; \; \ds \frac{1-p_{k,\ell}^{(b)}}{3} \\
	\end{array} \right.
\end{equation}
with $ p_{k,\ell}^{(b)}= \ds \frac{1}{1+C_{k,\ell}^{(b)}}$, and 
\begin{equation}
	C_{k,\ell}^{(b)}=\left \lbrace
	\begin{array}{ll}
		0 & \text{if} \; h_{k} < 13\; \text{m} \\
		 \left( \ds \frac{h_{k}-13\; \text{m}}{10}\right)^{1.5} g_{k,\ell}^{(b)}\; &  \text{if} \;   13\;\text{m}\leq h_{k} \leq 23\;\text{m}
	\end{array} \right. ,
\end{equation}
\begin{equation}
	g_{k,\ell}^{(b)}=\left \lbrace
	\begin{array}{ll}
		 0  & \text{if} \; d_{k,\ell, {\rm 2D}}^{(b)} \leq 18\; \text{m} \\
		\ds \frac{5}{4}\left( \ds \frac{d_{k,\ell}^{(b)}}{100}\right)^{3} \text{exp}\left( -\frac{d_{k,\ell}^{(b)}}{50}\right)   & \text{if} \;   d_{k,\ell}^{(b)} > 18\; \text{m}
	\end{array} \right. ,
\end{equation}
{
In the case of NLOS model the path-loss between the $k$-th user and the $\ell$-th BS is 
\begin{equation}
	\text{PL}_{k,\ell}^{(b)}=\max \left( \text{PL}_{k,\ell}^{(b), {\rm LOS}}, \widetilde{\text{PL}}_{k,\ell}^{(b)}\right) \, ,
	\label{PL_UMa_outdoor}
\end{equation}
where the term $ \widetilde{\text{PL}}_{k,\ell}^{(b)}$ is obtained as 
\begin{equation}
	\begin{array}{lll}
	\widetilde{\text{PL}}_{k,\ell}^{(b)} [\text{dB}]=& 13.54+39.08\log_{10} \left( d_{k,\ell, {\rm 3D}}^{(b)}\right) \\ & +20 \log_{10} (f_{c, {\rm GHz}}) -0.6\left( h_{k}-1.5\; \text{m}\right) \, .
	\end{array}
\end{equation}
The model in Eq. \eqref{PL_UMa_outdoor} holds for $1.5\; \text{m}<h_{k}<22.5\; \text{m}$ and $h_{\rm BS}=25$ m.}

\medskip

Let us now focus on the path loss model for the UE-AP channel. The path-loss between the $k$-th user and the $m$-th AP in the case of LOS is obtained as in Eq. \eqref{PL_dB_AP}, shown on the next page, where $d_{k,m, {\rm 3D}}^{(a)}$ and $d_{k,m}^{(a)}$ are the 3D and 2D distances between the $k$-th user and the $m$-th AP measured in meters, respectively, and $ h_{\rm AP} $ is the height of the AP in meters.
\begin{figure*}
\begin{equation}
	\text{PL}_{k,m}^{(a), {\rm LOS}} [\text{dB}]= \left\lbrace
	\begin{array}{llll}
		& 32.4 + 21\log_{10} \left( d_{k,m, {\rm 3D}}^{(a)} \right) + 20\log_{10} (f_{c, {\rm GHz}}) \; \text{if} \; 10\;\text{m} \leq   d_{k,m, {\rm 2D}}^{(a)} \leq  \widetilde{d}_{k,m, {\rm BP}}^{(a)} \\
		& 32.4+ 40\log_{10} \left( d_{k,m, {\rm 3D}}^{(a)} \right) + 20\log_{10} (f_{c, {\rm GHz}}) \\ &\quad - 9.5 \log_{10} \left[ \left( \widetilde{d}_{k,m, {\rm BP}}^{(a)} \right)^2 + \left( h_{\rm AP} - h_{{\rm MS},k}\right)^2 \right] \; \text{if} \;  d_{k,m, {\rm 2D}}^{(a)} \geq  5\;\text{km} \\
	\end{array} \right.
	\label{PL_dB_AP}
\end{equation}
\hrule
\hfill
\end{figure*}
The breakpoint distance $\widetilde{d}_{k,m, {\rm BP}}^{(a)}$ is obtained as
\begin{equation}
	\widetilde{d}_{k,m, {\rm BP}}^{(a)}= 4 \ds \frac{h_{\rm AP}^{'}}{h_{k}^{'}} \frac{f_c}{c},
\end{equation}
with $h_{\rm AP}^{'}$ and $h_{k}^{'}$ are the effective antenna heights at the AP and the $k$-th user, respectively. The effective antenna heights are calculated as $h_{\rm AP}^{'} = h_{\rm AP}-h_{k,m,{\rm E}}^{(a)}$, $h_{k}^{'} = h_{k}-h_{k,m,{\rm E}}^{(a)}$,
where $h_{k,m,{\rm E}}^{(a)}=1$m is the effective environment height. 
{In the case of NLOS, $\text{PL}_{k,m}^{(a)}$ is obtained as 
\begin{equation}
	\text{PL}_{k,m}^{(a)}=\max \left( \text{PL}_{k,m}^{(a), {\rm LOS}}, \widetilde{\text{PL}}_{k,m}^{(a)}\right) \, ,
	\label{PL_UMi_outdoor}
\end{equation}
where the term $ \widetilde{\text{PL}}_{k,m}^{(a)}$ is obtained as 
\begin{equation}
	\begin{array}{ll}
	\widetilde{\text{PL}}_{k,m}^{(a)} [\text{dB}]=& 22.4+35.3\log_{10} \left( d_{k,m, {\rm 3D}}^{(a)}\right) \\ & +21.3 \log_{10} (f_{c, {\rm GHz}}) -0.3\left( h_{k}-1.5\;\text{m}\right) \, .
		\end{array}
\end{equation}
The model in Eq. \eqref{PL_UMi_outdoor} holds for $1.5\;\text{m}<h_{k}<22.5\;\text{m}$ and $h_{\rm AP}=10$.}

\bibliographystyle{IEEEtran}
\bibliography{Cell_free_references}

\end{document}